\documentclass[usenatbib,usegraphicx]{mn2e}
\usepackage{comment}
\usepackage{amsfonts}
\usepackage{amsmath}
\usepackage{amssymb}
\usepackage{color}
\definecolor{AliceBlue}{rgb}{0.94,0.97,1.00}
\definecolor{AntiqueWhite1}{rgb}{1.00,0.94,0.86}
\definecolor{AntiqueWhite2}{rgb}{0.93,0.87,0.80}
\definecolor{AntiqueWhite3}{rgb}{0.80,0.75,0.69}
\definecolor{AntiqueWhite4}{rgb}{0.55,0.51,0.47}
\definecolor{AntiqueWhite}{rgb}{0.98,0.92,0.84}
\definecolor{BlanchedAlmond}{rgb}{1.00,0.92,0.80}
\definecolor{BlueViolet}{rgb}{0.54,0.17,0.89}
\definecolor{CadetBlue1}{rgb}{0.60,0.96,1.00}
\definecolor{CadetBlue2}{rgb}{0.56,0.90,0.93}
\definecolor{CadetBlue3}{rgb}{0.48,0.77,0.80}
\definecolor{CadetBlue4}{rgb}{0.33,0.53,0.55}
\definecolor{CadetBlue}{rgb}{0.37,0.62,0.63}
\definecolor{CornflowerBlue}{rgb}{0.39,0.58,0.93}
\definecolor{DarkBlue}{rgb}{0.00,0.00,0.55}
\definecolor{DarkCyan}{rgb}{0.00,0.55,0.55}
\definecolor{DarkGoldenrod1}{rgb}{1.00,0.73,0.06}
\definecolor{DarkGoldenrod2}{rgb}{0.93,0.68,0.05}
\definecolor{DarkGoldenrod3}{rgb}{0.80,0.58,0.05}
\definecolor{DarkGoldenrod4}{rgb}{0.55,0.40,0.03}
\definecolor{DarkGoldenrod}{rgb}{0.72,0.53,0.04}
\definecolor{DarkGray}{rgb}{0.66,0.66,0.66}
\definecolor{DarkGreen}{rgb}{0.00,0.39,0.00}
\definecolor{DarkGrey}{rgb}{0.66,0.66,0.66}
\definecolor{DarkKhaki}{rgb}{0.74,0.72,0.42}
\definecolor{DarkMagenta}{rgb}{0.55,0.00,0.55}
\definecolor{DarkOliveGreen1}{rgb}{0.79,1.00,0.44}
\definecolor{DarkOliveGreen2}{rgb}{0.74,0.93,0.41}
\definecolor{DarkOliveGreen3}{rgb}{0.64,0.80,0.35}
\definecolor{DarkOliveGreen4}{rgb}{0.43,0.55,0.24}
\definecolor{DarkOliveGreen}{rgb}{0.33,0.42,0.18}
\definecolor{DarkOrange1}{rgb}{1.00,0.50,0.00}
\definecolor{DarkOrange2}{rgb}{0.93,0.46,0.00}
\definecolor{DarkOrange3}{rgb}{0.80,0.40,0.00}
\definecolor{DarkOrange4}{rgb}{0.55,0.27,0.00}
\definecolor{DarkOrange}{rgb}{1.00,0.55,0.00}
\definecolor{DarkOrchid1}{rgb}{0.75,0.24,1.00}
\definecolor{DarkOrchid2}{rgb}{0.70,0.23,0.93}
\definecolor{DarkOrchid3}{rgb}{0.60,0.20,0.80}
\definecolor{DarkOrchid4}{rgb}{0.41,0.13,0.55}
\definecolor{DarkOrchid}{rgb}{0.60,0.20,0.80}
\definecolor{DarkRed}{rgb}{0.55,0.00,0.00}
\definecolor{DarkSalmon}{rgb}{0.91,0.59,0.48}
\definecolor{DarkSeaGreen1}{rgb}{0.76,1.00,0.76}
\definecolor{DarkSeaGreen2}{rgb}{0.71,0.93,0.71}
\definecolor{DarkSeaGreen3}{rgb}{0.61,0.80,0.61}
\definecolor{DarkSeaGreen4}{rgb}{0.41,0.55,0.41}
\definecolor{DarkSeaGreen}{rgb}{0.56,0.74,0.56}
\definecolor{DarkSlateBlue}{rgb}{0.28,0.24,0.55}
\definecolor{DarkSlateGray1}{rgb}{0.59,1.00,1.00}
\definecolor{DarkSlateGray2}{rgb}{0.55,0.93,0.93}
\definecolor{DarkSlateGray3}{rgb}{0.47,0.80,0.80}
\definecolor{DarkSlateGray4}{rgb}{0.32,0.55,0.55}
\definecolor{DarkSlateGray}{rgb}{0.18,0.31,0.31}
\definecolor{DarkSlateGrey}{rgb}{0.18,0.31,0.31}
\definecolor{DarkTurquoise}{rgb}{0.00,0.81,0.82}
\definecolor{DarkViolet}{rgb}{0.58,0.00,0.83}
\definecolor{DeepPink1}{rgb}{1.00,0.08,0.58}
\definecolor{DeepPink2}{rgb}{0.93,0.07,0.54}
\definecolor{DeepPink3}{rgb}{0.80,0.06,0.46}
\definecolor{DeepPink4}{rgb}{0.55,0.04,0.31}
\definecolor{DeepPink}{rgb}{1.00,0.08,0.58}
\definecolor{DeepSkyBlue1}{rgb}{0.00,0.75,1.00}
\definecolor{DeepSkyBlue2}{rgb}{0.00,0.70,0.93}
\definecolor{DeepSkyBlue3}{rgb}{0.00,0.60,0.80}
\definecolor{DeepSkyBlue4}{rgb}{0.00,0.41,0.55}
\definecolor{DeepSkyBlue}{rgb}{0.00,0.75,1.00}
\definecolor{DimGray}{rgb}{0.41,0.41,0.41}
\definecolor{DimGrey}{rgb}{0.41,0.41,0.41}
\definecolor{DodgerBlue1}{rgb}{0.12,0.56,1.00}
\definecolor{DodgerBlue2}{rgb}{0.11,0.53,0.93}
\definecolor{DodgerBlue3}{rgb}{0.09,0.45,0.80}
\definecolor{DodgerBlue4}{rgb}{0.06,0.31,0.55}
\definecolor{DodgerBlue}{rgb}{0.12,0.56,1.00}
\definecolor{FloralWhite}{rgb}{1.00,0.98,0.94}
\definecolor{ForestGreen}{rgb}{0.13,0.55,0.13}
\definecolor{GhostWhite}{rgb}{0.97,0.97,1.00}
\definecolor{GreenYellow}{rgb}{0.68,1.00,0.18}
\definecolor{HotPink1}{rgb}{1.00,0.43,0.71}
\definecolor{HotPink2}{rgb}{0.93,0.42,0.65}
\definecolor{HotPink3}{rgb}{0.80,0.38,0.56}
\definecolor{HotPink4}{rgb}{0.55,0.23,0.38}
\definecolor{HotPink}{rgb}{1.00,0.41,0.71}
\definecolor{IndianRed1}{rgb}{1.00,0.42,0.42}
\definecolor{IndianRed2}{rgb}{0.93,0.39,0.39}
\definecolor{IndianRed3}{rgb}{0.80,0.33,0.33}
\definecolor{IndianRed4}{rgb}{0.55,0.23,0.23}
\definecolor{IndianRed}{rgb}{0.80,0.36,0.36}
\definecolor{LavenderBlush1}{rgb}{1.00,0.94,0.96}
\definecolor{LavenderBlush2}{rgb}{0.93,0.88,0.90}
\definecolor{LavenderBlush3}{rgb}{0.80,0.76,0.77}
\definecolor{LavenderBlush4}{rgb}{0.55,0.51,0.53}
\definecolor{LavenderBlush}{rgb}{1.00,0.94,0.96}
\definecolor{LawnGreen}{rgb}{0.49,0.99,0.00}
\definecolor{LemonChiffon1}{rgb}{1.00,0.98,0.80}
\definecolor{LemonChiffon2}{rgb}{0.93,0.91,0.75}
\definecolor{LemonChiffon3}{rgb}{0.80,0.79,0.65}
\definecolor{LemonChiffon4}{rgb}{0.55,0.54,0.44}
\definecolor{LemonChiffon}{rgb}{1.00,0.98,0.80}
\definecolor{LightBlue1}{rgb}{0.75,0.94,1.00}
\definecolor{LightBlue2}{rgb}{0.70,0.87,0.93}
\definecolor{LightBlue3}{rgb}{0.60,0.75,0.80}
\definecolor{LightBlue4}{rgb}{0.41,0.51,0.55}
\definecolor{LightBlue}{rgb}{0.68,0.85,0.90}
\definecolor{LightCoral}{rgb}{0.94,0.50,0.50}
\definecolor{LightCyan1}{rgb}{0.88,1.00,1.00}
\definecolor{LightCyan2}{rgb}{0.82,0.93,0.93}
\definecolor{LightCyan3}{rgb}{0.71,0.80,0.80}
\definecolor{LightCyan4}{rgb}{0.48,0.55,0.55}
\definecolor{LightCyan}{rgb}{0.88,1.00,1.00}
\definecolor{LightGoldenrod1}{rgb}{1.00,0.93,0.55}
\definecolor{LightGoldenrod2}{rgb}{0.93,0.86,0.51}
\definecolor{LightGoldenrod3}{rgb}{0.80,0.75,0.44}
\definecolor{LightGoldenrod4}{rgb}{0.55,0.51,0.30}
\definecolor{LightGoldenrodYellow}{rgb}{0.98,0.98,0.82}
\definecolor{LightGoldenrod}{rgb}{0.93,0.87,0.51}
\definecolor{LightGray}{rgb}{0.83,0.83,0.83}
\definecolor{LightGreen}{rgb}{0.56,0.93,0.56}
\definecolor{LightGrey}{rgb}{0.83,0.83,0.83}
\definecolor{LightPink1}{rgb}{1.00,0.68,0.73}
\definecolor{LightPink2}{rgb}{0.93,0.64,0.68}
\definecolor{LightPink3}{rgb}{0.80,0.55,0.58}
\definecolor{LightPink4}{rgb}{0.55,0.37,0.40}
\definecolor{LightPink}{rgb}{1.00,0.71,0.76}
\definecolor{LightSalmon1}{rgb}{1.00,0.63,0.48}
\definecolor{LightSalmon2}{rgb}{0.93,0.58,0.45}
\definecolor{LightSalmon3}{rgb}{0.80,0.51,0.38}
\definecolor{LightSalmon4}{rgb}{0.55,0.34,0.26}
\definecolor{LightSalmon}{rgb}{1.00,0.63,0.48}
\definecolor{LightSeaGreen}{rgb}{0.13,0.70,0.67}
\definecolor{LightSkyBlue1}{rgb}{0.69,0.89,1.00}
\definecolor{LightSkyBlue2}{rgb}{0.64,0.83,0.93}
\definecolor{LightSkyBlue3}{rgb}{0.55,0.71,0.80}
\definecolor{LightSkyBlue4}{rgb}{0.38,0.48,0.55}
\definecolor{LightSkyBlue}{rgb}{0.53,0.81,0.98}
\definecolor{LightSlateBlue}{rgb}{0.52,0.44,1.00}
\definecolor{LightSlateGray}{rgb}{0.47,0.53,0.60}
\definecolor{LightSlateGrey}{rgb}{0.47,0.53,0.60}
\definecolor{LightSteelBlue1}{rgb}{0.79,0.88,1.00}
\definecolor{LightSteelBlue2}{rgb}{0.74,0.82,0.93}
\definecolor{LightSteelBlue3}{rgb}{0.64,0.71,0.80}
\definecolor{LightSteelBlue4}{rgb}{0.43,0.48,0.55}
\definecolor{LightSteelBlue}{rgb}{0.69,0.77,0.87}
\definecolor{LightYellow1}{rgb}{1.00,1.00,0.88}
\definecolor{LightYellow2}{rgb}{0.93,0.93,0.82}
\definecolor{LightYellow3}{rgb}{0.80,0.80,0.71}
\definecolor{LightYellow4}{rgb}{0.55,0.55,0.48}
\definecolor{LightYellow}{rgb}{1.00,1.00,0.88}
\definecolor{LimeGreen}{rgb}{0.20,0.80,0.20}
\definecolor{MediumAquamarine}{rgb}{0.40,0.80,0.67}
\definecolor{MediumBlue}{rgb}{0.00,0.00,0.80}
\definecolor{MediumOrchid1}{rgb}{0.88,0.40,1.00}
\definecolor{MediumOrchid2}{rgb}{0.82,0.37,0.93}
\definecolor{MediumOrchid3}{rgb}{0.71,0.32,0.80}
\definecolor{MediumOrchid4}{rgb}{0.48,0.22,0.55}
\definecolor{MediumOrchid}{rgb}{0.73,0.33,0.83}
\definecolor{MediumPurple1}{rgb}{0.67,0.51,1.00}
\definecolor{MediumPurple2}{rgb}{0.62,0.47,0.93}
\definecolor{MediumPurple3}{rgb}{0.54,0.41,0.80}
\definecolor{MediumPurple4}{rgb}{0.36,0.28,0.55}
\definecolor{MediumPurple}{rgb}{0.58,0.44,0.86}
\definecolor{MediumSeaGreen}{rgb}{0.24,0.70,0.44}
\definecolor{MediumSlateBlue}{rgb}{0.48,0.41,0.93}
\definecolor{MediumSpringGreen}{rgb}{0.00,0.98,0.60}
\definecolor{MediumTurquoise}{rgb}{0.28,0.82,0.80}
\definecolor{MediumVioletRed}{rgb}{0.78,0.08,0.52}
\definecolor{MidnightBlue}{rgb}{0.10,0.10,0.44}
\definecolor{MintCream}{rgb}{0.96,1.00,0.98}
\definecolor{MistyRose1}{rgb}{1.00,0.89,0.88}
\definecolor{MistyRose2}{rgb}{0.93,0.84,0.82}
\definecolor{MistyRose3}{rgb}{0.80,0.72,0.71}
\definecolor{MistyRose4}{rgb}{0.55,0.49,0.48}
\definecolor{MistyRose}{rgb}{1.00,0.89,0.88}
\definecolor{NavajoWhite1}{rgb}{1.00,0.87,0.68}
\definecolor{NavajoWhite2}{rgb}{0.93,0.81,0.63}
\definecolor{NavajoWhite3}{rgb}{0.80,0.70,0.55}
\definecolor{NavajoWhite4}{rgb}{0.55,0.47,0.37}
\definecolor{NavajoWhite}{rgb}{1.00,0.87,0.68}
\definecolor{NavyBlue}{rgb}{0.00,0.00,0.50}
\definecolor{OldLace}{rgb}{0.99,0.96,0.90}
\definecolor{OliveDrab1}{rgb}{0.75,1.00,0.24}
\definecolor{OliveDrab2}{rgb}{0.70,0.93,0.23}
\definecolor{OliveDrab3}{rgb}{0.60,0.80,0.20}
\definecolor{OliveDrab4}{rgb}{0.41,0.55,0.13}
\definecolor{OliveDrab}{rgb}{0.42,0.56,0.14}
\definecolor{OrangeRed1}{rgb}{1.00,0.27,0.00}
\definecolor{OrangeRed2}{rgb}{0.93,0.25,0.00}
\definecolor{OrangeRed3}{rgb}{0.80,0.22,0.00}
\definecolor{OrangeRed4}{rgb}{0.55,0.15,0.00}
\definecolor{OrangeRed}{rgb}{1.00,0.27,0.00}
\definecolor{PaleGoldenrod}{rgb}{0.93,0.91,0.67}
\definecolor{PaleGreen1}{rgb}{0.60,1.00,0.60}
\definecolor{PaleGreen2}{rgb}{0.56,0.93,0.56}
\definecolor{PaleGreen3}{rgb}{0.49,0.80,0.49}
\definecolor{PaleGreen4}{rgb}{0.33,0.55,0.33}
\definecolor{PaleGreen}{rgb}{0.60,0.98,0.60}
\definecolor{PaleTurquoise1}{rgb}{0.73,1.00,1.00}
\definecolor{PaleTurquoise2}{rgb}{0.68,0.93,0.93}
\definecolor{PaleTurquoise3}{rgb}{0.59,0.80,0.80}
\definecolor{PaleTurquoise4}{rgb}{0.40,0.55,0.55}
\definecolor{PaleTurquoise}{rgb}{0.69,0.93,0.93}
\definecolor{PaleVioletRed1}{rgb}{1.00,0.51,0.67}
\definecolor{PaleVioletRed2}{rgb}{0.93,0.47,0.62}
\definecolor{PaleVioletRed3}{rgb}{0.80,0.41,0.54}
\definecolor{PaleVioletRed4}{rgb}{0.55,0.28,0.36}
\definecolor{PaleVioletRed}{rgb}{0.86,0.44,0.58}
\definecolor{PapayaWhip}{rgb}{1.00,0.94,0.84}
\definecolor{PeachPuff1}{rgb}{1.00,0.85,0.73}
\definecolor{PeachPuff2}{rgb}{0.93,0.80,0.68}
\definecolor{PeachPuff3}{rgb}{0.80,0.69,0.58}
\definecolor{PeachPuff4}{rgb}{0.55,0.47,0.40}
\definecolor{PeachPuff}{rgb}{1.00,0.85,0.73}
\definecolor{PowderBlue}{rgb}{0.69,0.88,0.90}
\definecolor{RosyBrown1}{rgb}{1.00,0.76,0.76}
\definecolor{RosyBrown2}{rgb}{0.93,0.71,0.71}
\definecolor{RosyBrown3}{rgb}{0.80,0.61,0.61}
\definecolor{RosyBrown4}{rgb}{0.55,0.41,0.41}
\definecolor{RosyBrown}{rgb}{0.74,0.56,0.56}
\definecolor{RoyalBlue1}{rgb}{0.28,0.46,1.00}
\definecolor{RoyalBlue2}{rgb}{0.26,0.43,0.93}
\definecolor{RoyalBlue3}{rgb}{0.23,0.37,0.80}
\definecolor{RoyalBlue4}{rgb}{0.15,0.25,0.55}
\definecolor{RoyalBlue}{rgb}{0.25,0.41,0.88}
\definecolor{SaddleBrown}{rgb}{0.55,0.27,0.07}
\definecolor{SandyBrown}{rgb}{0.96,0.64,0.38}
\definecolor{SeaGreen1}{rgb}{0.33,1.00,0.62}
\definecolor{SeaGreen2}{rgb}{0.31,0.93,0.58}
\definecolor{SeaGreen3}{rgb}{0.26,0.80,0.50}
\definecolor{SeaGreen4}{rgb}{0.18,0.55,0.34}
\definecolor{SeaGreen}{rgb}{0.18,0.55,0.34}
\definecolor{SkyBlue1}{rgb}{0.53,0.81,1.00}
\definecolor{SkyBlue2}{rgb}{0.49,0.75,0.93}
\definecolor{SkyBlue3}{rgb}{0.42,0.65,0.80}
\definecolor{SkyBlue4}{rgb}{0.29,0.44,0.55}
\definecolor{SkyBlue}{rgb}{0.53,0.81,0.92}
\definecolor{SlateBlue1}{rgb}{0.51,0.44,1.00}
\definecolor{SlateBlue2}{rgb}{0.48,0.40,0.93}
\definecolor{SlateBlue3}{rgb}{0.41,0.35,0.80}
\definecolor{SlateBlue4}{rgb}{0.28,0.24,0.55}
\definecolor{SlateBlue}{rgb}{0.42,0.35,0.80}
\definecolor{SlateGray1}{rgb}{0.78,0.89,1.00}
\definecolor{SlateGray2}{rgb}{0.73,0.83,0.93}
\definecolor{SlateGray3}{rgb}{0.62,0.71,0.80}
\definecolor{SlateGray4}{rgb}{0.42,0.48,0.55}
\definecolor{SlateGray}{rgb}{0.44,0.50,0.56}
\definecolor{SlateGrey}{rgb}{0.44,0.50,0.56}
\definecolor{SpringGreen1}{rgb}{0.00,1.00,0.50}
\definecolor{SpringGreen2}{rgb}{0.00,0.93,0.46}
\definecolor{SpringGreen3}{rgb}{0.00,0.80,0.40}
\definecolor{SpringGreen4}{rgb}{0.00,0.55,0.27}
\definecolor{SpringGreen}{rgb}{0.00,1.00,0.50}
\definecolor{SteelBlue1}{rgb}{0.39,0.72,1.00}
\definecolor{SteelBlue2}{rgb}{0.36,0.67,0.93}
\definecolor{SteelBlue3}{rgb}{0.31,0.58,0.80}
\definecolor{SteelBlue4}{rgb}{0.21,0.39,0.55}
\definecolor{SteelBlue}{rgb}{0.27,0.51,0.71}
\definecolor{VioletRed1}{rgb}{1.00,0.24,0.59}
\definecolor{VioletRed2}{rgb}{0.93,0.23,0.55}
\definecolor{VioletRed3}{rgb}{0.80,0.20,0.47}
\definecolor{VioletRed4}{rgb}{0.55,0.13,0.32}
\definecolor{VioletRed}{rgb}{0.82,0.13,0.56}
\definecolor{WhiteSmoke}{rgb}{0.96,0.96,0.96}
\definecolor{YellowGreen}{rgb}{0.60,0.80,0.20}
\definecolor{aliceblue}{rgb}{0.94,0.97,1.00}
\definecolor{antiquewhite}{rgb}{0.98,0.92,0.84}
\definecolor{aquamarine1}{rgb}{0.50,1.00,0.83}
\definecolor{aquamarine2}{rgb}{0.46,0.93,0.78}
\definecolor{aquamarine3}{rgb}{0.40,0.80,0.67}
\definecolor{aquamarine4}{rgb}{0.27,0.55,0.45}
\definecolor{aquamarine}{rgb}{0.50,1.00,0.83}
\definecolor{azure1}{rgb}{0.94,1.00,1.00}
\definecolor{azure2}{rgb}{0.88,0.93,0.93}
\definecolor{azure3}{rgb}{0.76,0.80,0.80}
\definecolor{azure4}{rgb}{0.51,0.55,0.55}
\definecolor{azure}{rgb}{0.94,1.00,1.00}
\definecolor{beige}{rgb}{0.96,0.96,0.86}
\definecolor{bisque1}{rgb}{1.00,0.89,0.77}
\definecolor{bisque2}{rgb}{0.93,0.84,0.72}
\definecolor{bisque3}{rgb}{0.80,0.72,0.62}
\definecolor{bisque4}{rgb}{0.55,0.49,0.42}
\definecolor{bisque}{rgb}{1.00,0.89,0.77}
\definecolor{black}{rgb}{0.00,0.00,0.00}
\definecolor{blanchedalmond}{rgb}{1.00,0.92,0.80}
\definecolor{blue1}{rgb}{0.00,0.00,1.00}
\definecolor{blue2}{rgb}{0.00,0.00,0.93}
\definecolor{blue3}{rgb}{0.00,0.00,0.80}
\definecolor{blue4}{rgb}{0.00,0.00,0.55}
\definecolor{blueviolet}{rgb}{0.54,0.17,0.89}
\definecolor{blue}{rgb}{0.00,0.00,1.00}
\definecolor{brown1}{rgb}{1.00,0.25,0.25}
\definecolor{brown2}{rgb}{0.93,0.23,0.23}
\definecolor{brown3}{rgb}{0.80,0.20,0.20}
\definecolor{brown4}{rgb}{0.55,0.14,0.14}
\definecolor{brown}{rgb}{0.65,0.16,0.16}
\definecolor{burlywood1}{rgb}{1.00,0.83,0.61}
\definecolor{burlywood2}{rgb}{0.93,0.77,0.57}
\definecolor{burlywood3}{rgb}{0.80,0.67,0.49}
\definecolor{burlywood4}{rgb}{0.55,0.45,0.33}
\definecolor{burlywood}{rgb}{0.87,0.72,0.53}
\definecolor{cadetblue}{rgb}{0.37,0.62,0.63}
\definecolor{chartreuse1}{rgb}{0.50,1.00,0.00}
\definecolor{chartreuse2}{rgb}{0.46,0.93,0.00}
\definecolor{chartreuse3}{rgb}{0.40,0.80,0.00}
\definecolor{chartreuse4}{rgb}{0.27,0.55,0.00}
\definecolor{chartreuse}{rgb}{0.50,1.00,0.00}
\definecolor{chocolate1}{rgb}{1.00,0.50,0.14}
\definecolor{chocolate2}{rgb}{0.93,0.46,0.13}
\definecolor{chocolate3}{rgb}{0.80,0.40,0.11}
\definecolor{chocolate4}{rgb}{0.55,0.27,0.07}
\definecolor{chocolate}{rgb}{0.82,0.41,0.12}
\definecolor{coral1}{rgb}{1.00,0.45,0.34}
\definecolor{coral2}{rgb}{0.93,0.42,0.31}
\definecolor{coral3}{rgb}{0.80,0.36,0.27}
\definecolor{coral4}{rgb}{0.55,0.24,0.18}
\definecolor{coral}{rgb}{1.00,0.50,0.31}
\definecolor{cornflowerblue}{rgb}{0.39,0.58,0.93}
\definecolor{cornsilk1}{rgb}{1.00,0.97,0.86}
\definecolor{cornsilk2}{rgb}{0.93,0.91,0.80}
\definecolor{cornsilk3}{rgb}{0.80,0.78,0.69}
\definecolor{cornsilk4}{rgb}{0.55,0.53,0.47}
\definecolor{cornsilk}{rgb}{1.00,0.97,0.86}
\definecolor{cyan1}{rgb}{0.00,1.00,1.00}
\definecolor{cyan2}{rgb}{0.00,0.93,0.93}
\definecolor{cyan3}{rgb}{0.00,0.80,0.80}
\definecolor{cyan4}{rgb}{0.00,0.55,0.55}
\definecolor{cyan}{rgb}{0.00,1.00,1.00}
\definecolor{darkblue}{rgb}{0.00,0.00,0.55}
\definecolor{darkcyan}{rgb}{0.00,0.55,0.55}
\definecolor{darkgoldenrod}{rgb}{0.72,0.53,0.04}
\definecolor{darkgray}{rgb}{0.66,0.66,0.66}
\definecolor{darkgreen}{rgb}{0.00,0.39,0.00}
\definecolor{darkgrey}{rgb}{0.66,0.66,0.66}
\definecolor{darkkhaki}{rgb}{0.74,0.72,0.42}
\definecolor{darkmagenta}{rgb}{0.55,0.00,0.55}
\definecolor{darkolive}{rgb}{0.33,0.42,0.18}
\definecolor{darkorange}{rgb}{1.00,0.55,0.00}
\definecolor{darkorchid}{rgb}{0.60,0.20,0.80}
\definecolor{darkred}{rgb}{0.55,0.00,0.00}
\definecolor{darksalmon}{rgb}{0.91,0.59,0.48}
\definecolor{darksea}{rgb}{0.56,0.74,0.56}
\definecolor{darkslate}{rgb}{0.18,0.31,0.31}
\definecolor{darkslate}{rgb}{0.18,0.31,0.31}
\definecolor{darkslate}{rgb}{0.28,0.24,0.55}
\definecolor{darkturquoise}{rgb}{0.00,0.81,0.82}
\definecolor{darkviolet}{rgb}{0.58,0.00,0.83}
\definecolor{deeppink}{rgb}{1.00,0.08,0.58}
\definecolor{deepsky}{rgb}{0.00,0.75,1.00}
\definecolor{dimgray}{rgb}{0.41,0.41,0.41}
\definecolor{dimgrey}{rgb}{0.41,0.41,0.41}
\definecolor{dodgerblue}{rgb}{0.12,0.56,1.00}
\definecolor{firebrick1}{rgb}{1.00,0.19,0.19}
\definecolor{firebrick2}{rgb}{0.93,0.17,0.17}
\definecolor{firebrick3}{rgb}{0.80,0.15,0.15}
\definecolor{firebrick4}{rgb}{0.55,0.10,0.10}
\definecolor{firebrick}{rgb}{0.70,0.13,0.13}
\definecolor{floralwhite}{rgb}{1.00,0.98,0.94}
\definecolor{forestgreen}{rgb}{0.13,0.55,0.13}
\definecolor{gainsboro}{rgb}{0.86,0.86,0.86}
\definecolor{ghostwhite}{rgb}{0.97,0.97,1.00}
\definecolor{gold1}{rgb}{1.00,0.84,0.00}
\definecolor{gold2}{rgb}{0.93,0.79,0.00}
\definecolor{gold3}{rgb}{0.80,0.68,0.00}
\definecolor{gold4}{rgb}{0.55,0.46,0.00}
\definecolor{goldenrod1}{rgb}{1.00,0.76,0.15}
\definecolor{goldenrod2}{rgb}{0.93,0.71,0.13}
\definecolor{goldenrod3}{rgb}{0.80,0.61,0.11}
\definecolor{goldenrod4}{rgb}{0.55,0.41,0.08}
\definecolor{goldenrod}{rgb}{0.85,0.65,0.13}
\definecolor{gold}{rgb}{1.00,0.84,0.00}
\definecolor{gray0}{rgb}{0.00,0.00,0.00}
\definecolor{gray100}{rgb}{1.00,1.00,1.00}
\definecolor{gray10}{rgb}{0.10,0.10,0.10}
\definecolor{gray11}{rgb}{0.11,0.11,0.11}
\definecolor{gray12}{rgb}{0.12,0.12,0.12}
\definecolor{gray13}{rgb}{0.13,0.13,0.13}
\definecolor{gray14}{rgb}{0.14,0.14,0.14}
\definecolor{gray15}{rgb}{0.15,0.15,0.15}
\definecolor{gray16}{rgb}{0.16,0.16,0.16}
\definecolor{gray17}{rgb}{0.17,0.17,0.17}
\definecolor{gray18}{rgb}{0.18,0.18,0.18}
\definecolor{gray19}{rgb}{0.19,0.19,0.19}
\definecolor{gray1}{rgb}{0.01,0.01,0.01}
\definecolor{gray20}{rgb}{0.20,0.20,0.20}
\definecolor{gray21}{rgb}{0.21,0.21,0.21}
\definecolor{gray22}{rgb}{0.22,0.22,0.22}
\definecolor{gray23}{rgb}{0.23,0.23,0.23}
\definecolor{gray24}{rgb}{0.24,0.24,0.24}
\definecolor{gray25}{rgb}{0.25,0.25,0.25}
\definecolor{gray26}{rgb}{0.26,0.26,0.26}
\definecolor{gray27}{rgb}{0.27,0.27,0.27}
\definecolor{gray28}{rgb}{0.28,0.28,0.28}
\definecolor{gray29}{rgb}{0.29,0.29,0.29}
\definecolor{gray2}{rgb}{0.02,0.02,0.02}
\definecolor{gray30}{rgb}{0.30,0.30,0.30}
\definecolor{gray31}{rgb}{0.31,0.31,0.31}
\definecolor{gray32}{rgb}{0.32,0.32,0.32}
\definecolor{gray33}{rgb}{0.33,0.33,0.33}
\definecolor{gray34}{rgb}{0.34,0.34,0.34}
\definecolor{gray35}{rgb}{0.35,0.35,0.35}
\definecolor{gray36}{rgb}{0.36,0.36,0.36}
\definecolor{gray37}{rgb}{0.37,0.37,0.37}
\definecolor{gray38}{rgb}{0.38,0.38,0.38}
\definecolor{gray39}{rgb}{0.39,0.39,0.39}
\definecolor{gray3}{rgb}{0.03,0.03,0.03}
\definecolor{gray40}{rgb}{0.40,0.40,0.40}
\definecolor{gray41}{rgb}{0.41,0.41,0.41}
\definecolor{gray42}{rgb}{0.42,0.42,0.42}
\definecolor{gray43}{rgb}{0.43,0.43,0.43}
\definecolor{gray44}{rgb}{0.44,0.44,0.44}
\definecolor{gray45}{rgb}{0.45,0.45,0.45}
\definecolor{gray46}{rgb}{0.46,0.46,0.46}
\definecolor{gray47}{rgb}{0.47,0.47,0.47}
\definecolor{gray48}{rgb}{0.48,0.48,0.48}
\definecolor{gray49}{rgb}{0.49,0.49,0.49}
\definecolor{gray4}{rgb}{0.04,0.04,0.04}
\definecolor{gray50}{rgb}{0.50,0.50,0.50}
\definecolor{gray51}{rgb}{0.51,0.51,0.51}
\definecolor{gray52}{rgb}{0.52,0.52,0.52}
\definecolor{gray53}{rgb}{0.53,0.53,0.53}
\definecolor{gray54}{rgb}{0.54,0.54,0.54}
\definecolor{gray55}{rgb}{0.55,0.55,0.55}
\definecolor{gray56}{rgb}{0.56,0.56,0.56}
\definecolor{gray57}{rgb}{0.57,0.57,0.57}
\definecolor{gray58}{rgb}{0.58,0.58,0.58}
\definecolor{gray59}{rgb}{0.59,0.59,0.59}
\definecolor{gray5}{rgb}{0.05,0.05,0.05}
\definecolor{gray60}{rgb}{0.60,0.60,0.60}
\definecolor{gray61}{rgb}{0.61,0.61,0.61}
\definecolor{gray62}{rgb}{0.62,0.62,0.62}
\definecolor{gray63}{rgb}{0.63,0.63,0.63}
\definecolor{gray64}{rgb}{0.64,0.64,0.64}
\definecolor{gray65}{rgb}{0.65,0.65,0.65}
\definecolor{gray66}{rgb}{0.66,0.66,0.66}
\definecolor{gray67}{rgb}{0.67,0.67,0.67}
\definecolor{gray68}{rgb}{0.68,0.68,0.68}
\definecolor{gray69}{rgb}{0.69,0.69,0.69}
\definecolor{gray6}{rgb}{0.06,0.06,0.06}
\definecolor{gray70}{rgb}{0.70,0.70,0.70}
\definecolor{gray71}{rgb}{0.71,0.71,0.71}
\definecolor{gray72}{rgb}{0.72,0.72,0.72}
\definecolor{gray73}{rgb}{0.73,0.73,0.73}
\definecolor{gray74}{rgb}{0.74,0.74,0.74}
\definecolor{gray75}{rgb}{0.75,0.75,0.75}
\definecolor{gray76}{rgb}{0.76,0.76,0.76}
\definecolor{gray77}{rgb}{0.77,0.77,0.77}
\definecolor{gray78}{rgb}{0.78,0.78,0.78}
\definecolor{gray79}{rgb}{0.79,0.79,0.79}
\definecolor{gray7}{rgb}{0.07,0.07,0.07}
\definecolor{gray80}{rgb}{0.80,0.80,0.80}
\definecolor{gray81}{rgb}{0.81,0.81,0.81}
\definecolor{gray82}{rgb}{0.82,0.82,0.82}
\definecolor{gray83}{rgb}{0.83,0.83,0.83}
\definecolor{gray84}{rgb}{0.84,0.84,0.84}
\definecolor{gray85}{rgb}{0.85,0.85,0.85}
\definecolor{gray86}{rgb}{0.86,0.86,0.86}
\definecolor{gray87}{rgb}{0.87,0.87,0.87}
\definecolor{gray88}{rgb}{0.88,0.88,0.88}
\definecolor{gray89}{rgb}{0.89,0.89,0.89}
\definecolor{gray8}{rgb}{0.08,0.08,0.08}
\definecolor{gray90}{rgb}{0.90,0.90,0.90}
\definecolor{gray91}{rgb}{0.91,0.91,0.91}
\definecolor{gray92}{rgb}{0.92,0.92,0.92}
\definecolor{gray93}{rgb}{0.93,0.93,0.93}
\definecolor{gray94}{rgb}{0.94,0.94,0.94}
\definecolor{gray95}{rgb}{0.95,0.95,0.95}
\definecolor{gray96}{rgb}{0.96,0.96,0.96}
\definecolor{gray97}{rgb}{0.97,0.97,0.97}
\definecolor{gray98}{rgb}{0.98,0.98,0.98}
\definecolor{gray99}{rgb}{0.99,0.99,0.99}
\definecolor{gray9}{rgb}{0.09,0.09,0.09}
\definecolor{gray}{rgb}{0.75,0.75,0.75}
\definecolor{green1}{rgb}{0.00,1.00,0.00}
\definecolor{green2}{rgb}{0.00,0.93,0.00}
\definecolor{green3}{rgb}{0.00,0.80,0.00}
\definecolor{green4}{rgb}{0.00,0.55,0.00}
\definecolor{greenyellow}{rgb}{0.68,1.00,0.18}
\definecolor{green}{rgb}{0.00,1.00,0.00}
\definecolor{grey0}{rgb}{0.00,0.00,0.00}
\definecolor{grey100}{rgb}{1.00,1.00,1.00}
\definecolor{grey10}{rgb}{0.10,0.10,0.10}
\definecolor{grey11}{rgb}{0.11,0.11,0.11}
\definecolor{grey12}{rgb}{0.12,0.12,0.12}
\definecolor{grey13}{rgb}{0.13,0.13,0.13}
\definecolor{grey14}{rgb}{0.14,0.14,0.14}
\definecolor{grey15}{rgb}{0.15,0.15,0.15}
\definecolor{grey16}{rgb}{0.16,0.16,0.16}
\definecolor{grey17}{rgb}{0.17,0.17,0.17}
\definecolor{grey18}{rgb}{0.18,0.18,0.18}
\definecolor{grey19}{rgb}{0.19,0.19,0.19}
\definecolor{grey1}{rgb}{0.01,0.01,0.01}
\definecolor{grey20}{rgb}{0.20,0.20,0.20}
\definecolor{grey21}{rgb}{0.21,0.21,0.21}
\definecolor{grey22}{rgb}{0.22,0.22,0.22}
\definecolor{grey23}{rgb}{0.23,0.23,0.23}
\definecolor{grey24}{rgb}{0.24,0.24,0.24}
\definecolor{grey25}{rgb}{0.25,0.25,0.25}
\definecolor{grey26}{rgb}{0.26,0.26,0.26}
\definecolor{grey27}{rgb}{0.27,0.27,0.27}
\definecolor{grey28}{rgb}{0.28,0.28,0.28}
\definecolor{grey29}{rgb}{0.29,0.29,0.29}
\definecolor{grey2}{rgb}{0.02,0.02,0.02}
\definecolor{grey30}{rgb}{0.30,0.30,0.30}
\definecolor{grey31}{rgb}{0.31,0.31,0.31}
\definecolor{grey32}{rgb}{0.32,0.32,0.32}
\definecolor{grey33}{rgb}{0.33,0.33,0.33}
\definecolor{grey34}{rgb}{0.34,0.34,0.34}
\definecolor{grey35}{rgb}{0.35,0.35,0.35}
\definecolor{grey36}{rgb}{0.36,0.36,0.36}
\definecolor{grey37}{rgb}{0.37,0.37,0.37}
\definecolor{grey38}{rgb}{0.38,0.38,0.38}
\definecolor{grey39}{rgb}{0.39,0.39,0.39}
\definecolor{grey3}{rgb}{0.03,0.03,0.03}
\definecolor{grey40}{rgb}{0.40,0.40,0.40}
\definecolor{grey41}{rgb}{0.41,0.41,0.41}
\definecolor{grey42}{rgb}{0.42,0.42,0.42}
\definecolor{grey43}{rgb}{0.43,0.43,0.43}
\definecolor{grey44}{rgb}{0.44,0.44,0.44}
\definecolor{grey45}{rgb}{0.45,0.45,0.45}
\definecolor{grey46}{rgb}{0.46,0.46,0.46}
\definecolor{grey47}{rgb}{0.47,0.47,0.47}
\definecolor{grey48}{rgb}{0.48,0.48,0.48}
\definecolor{grey49}{rgb}{0.49,0.49,0.49}
\definecolor{grey4}{rgb}{0.04,0.04,0.04}
\definecolor{grey50}{rgb}{0.50,0.50,0.50}
\definecolor{grey51}{rgb}{0.51,0.51,0.51}
\definecolor{grey52}{rgb}{0.52,0.52,0.52}
\definecolor{grey53}{rgb}{0.53,0.53,0.53}
\definecolor{grey54}{rgb}{0.54,0.54,0.54}
\definecolor{grey55}{rgb}{0.55,0.55,0.55}
\definecolor{grey56}{rgb}{0.56,0.56,0.56}
\definecolor{grey57}{rgb}{0.57,0.57,0.57}
\definecolor{grey58}{rgb}{0.58,0.58,0.58}
\definecolor{grey59}{rgb}{0.59,0.59,0.59}
\definecolor{grey5}{rgb}{0.05,0.05,0.05}
\definecolor{grey60}{rgb}{0.60,0.60,0.60}
\definecolor{grey61}{rgb}{0.61,0.61,0.61}
\definecolor{grey62}{rgb}{0.62,0.62,0.62}
\definecolor{grey63}{rgb}{0.63,0.63,0.63}
\definecolor{grey64}{rgb}{0.64,0.64,0.64}
\definecolor{grey65}{rgb}{0.65,0.65,0.65}
\definecolor{grey66}{rgb}{0.66,0.66,0.66}
\definecolor{grey67}{rgb}{0.67,0.67,0.67}
\definecolor{grey68}{rgb}{0.68,0.68,0.68}
\definecolor{grey69}{rgb}{0.69,0.69,0.69}
\definecolor{grey6}{rgb}{0.06,0.06,0.06}
\definecolor{grey70}{rgb}{0.70,0.70,0.70}
\definecolor{grey71}{rgb}{0.71,0.71,0.71}
\definecolor{grey72}{rgb}{0.72,0.72,0.72}
\definecolor{grey73}{rgb}{0.73,0.73,0.73}
\definecolor{grey74}{rgb}{0.74,0.74,0.74}
\definecolor{grey75}{rgb}{0.75,0.75,0.75}
\definecolor{grey76}{rgb}{0.76,0.76,0.76}
\definecolor{grey77}{rgb}{0.77,0.77,0.77}
\definecolor{grey78}{rgb}{0.78,0.78,0.78}
\definecolor{grey79}{rgb}{0.79,0.79,0.79}
\definecolor{grey7}{rgb}{0.07,0.07,0.07}
\definecolor{grey80}{rgb}{0.80,0.80,0.80}
\definecolor{grey81}{rgb}{0.81,0.81,0.81}
\definecolor{grey82}{rgb}{0.82,0.82,0.82}
\definecolor{grey83}{rgb}{0.83,0.83,0.83}
\definecolor{grey84}{rgb}{0.84,0.84,0.84}
\definecolor{grey85}{rgb}{0.85,0.85,0.85}
\definecolor{grey86}{rgb}{0.86,0.86,0.86}
\definecolor{grey87}{rgb}{0.87,0.87,0.87}
\definecolor{grey88}{rgb}{0.88,0.88,0.88}
\definecolor{grey89}{rgb}{0.89,0.89,0.89}
\definecolor{grey8}{rgb}{0.08,0.08,0.08}
\definecolor{grey90}{rgb}{0.90,0.90,0.90}
\definecolor{grey91}{rgb}{0.91,0.91,0.91}
\definecolor{grey92}{rgb}{0.92,0.92,0.92}
\definecolor{grey93}{rgb}{0.93,0.93,0.93}
\definecolor{grey94}{rgb}{0.94,0.94,0.94}
\definecolor{grey95}{rgb}{0.95,0.95,0.95}
\definecolor{grey96}{rgb}{0.96,0.96,0.96}
\definecolor{grey97}{rgb}{0.97,0.97,0.97}
\definecolor{grey98}{rgb}{0.98,0.98,0.98}
\definecolor{grey99}{rgb}{0.99,0.99,0.99}
\definecolor{grey9}{rgb}{0.09,0.09,0.09}
\definecolor{grey}{rgb}{0.75,0.75,0.75}
\definecolor{honeydew1}{rgb}{0.94,1.00,0.94}
\definecolor{honeydew2}{rgb}{0.88,0.93,0.88}
\definecolor{honeydew3}{rgb}{0.76,0.80,0.76}
\definecolor{honeydew4}{rgb}{0.51,0.55,0.51}
\definecolor{honeydew}{rgb}{0.94,1.00,0.94}
\definecolor{hotpink}{rgb}{1.00,0.41,0.71}
\definecolor{indianred}{rgb}{0.80,0.36,0.36}
\definecolor{ivory1}{rgb}{1.00,1.00,0.94}
\definecolor{ivory2}{rgb}{0.93,0.93,0.88}
\definecolor{ivory3}{rgb}{0.80,0.80,0.76}
\definecolor{ivory4}{rgb}{0.55,0.55,0.51}
\definecolor{ivory}{rgb}{1.00,1.00,0.94}
\definecolor{khaki1}{rgb}{1.00,0.96,0.56}
\definecolor{khaki2}{rgb}{0.93,0.90,0.52}
\definecolor{khaki3}{rgb}{0.80,0.78,0.45}
\definecolor{khaki4}{rgb}{0.55,0.53,0.31}
\definecolor{khaki}{rgb}{0.94,0.90,0.55}
\definecolor{lavenderblush}{rgb}{1.00,0.94,0.96}
\definecolor{lavender}{rgb}{0.90,0.90,0.98}
\definecolor{lawngreen}{rgb}{0.49,0.99,0.00}
\definecolor{lemonchiffon}{rgb}{1.00,0.98,0.80}
\definecolor{lightblue}{rgb}{0.68,0.85,0.90}
\definecolor{lightcoral}{rgb}{0.94,0.50,0.50}
\definecolor{lightcyan}{rgb}{0.88,1.00,1.00}
\definecolor{lightgoldenrod}{rgb}{0.93,0.87,0.51}
\definecolor{lightgoldenrod}{rgb}{0.98,0.98,0.82}
\definecolor{lightgray}{rgb}{0.83,0.83,0.83}
\definecolor{lightgreen}{rgb}{0.56,0.93,0.56}
\definecolor{lightgrey}{rgb}{0.83,0.83,0.83}
\definecolor{lightpink}{rgb}{1.00,0.71,0.76}
\definecolor{lightsalmon}{rgb}{1.00,0.63,0.48}
\definecolor{lightsea}{rgb}{0.13,0.70,0.67}
\definecolor{lightsky}{rgb}{0.53,0.81,0.98}
\definecolor{lightslate}{rgb}{0.47,0.53,0.60}
\definecolor{lightslate}{rgb}{0.47,0.53,0.60}
\definecolor{lightslate}{rgb}{0.52,0.44,1.00}
\definecolor{lightsteel}{rgb}{0.69,0.77,0.87}
\definecolor{lightyellow}{rgb}{1.00,1.00,0.88}
\definecolor{limegreen}{rgb}{0.20,0.80,0.20}
\definecolor{linen}{rgb}{0.98,0.94,0.90}
\definecolor{magenta1}{rgb}{1.00,0.00,1.00}
\definecolor{magenta2}{rgb}{0.93,0.00,0.93}
\definecolor{magenta3}{rgb}{0.80,0.00,0.80}
\definecolor{magenta4}{rgb}{0.55,0.00,0.55}
\definecolor{magenta}{rgb}{1.00,0.00,1.00}
\definecolor{maroon1}{rgb}{1.00,0.20,0.70}
\definecolor{maroon2}{rgb}{0.93,0.19,0.65}
\definecolor{maroon3}{rgb}{0.80,0.16,0.56}
\definecolor{maroon4}{rgb}{0.55,0.11,0.38}
\definecolor{maroon}{rgb}{0.69,0.19,0.38}
\definecolor{mediumaquamarine}{rgb}{0.40,0.80,0.67}
\definecolor{mediumblue}{rgb}{0.00,0.00,0.80}
\definecolor{mediumorchid}{rgb}{0.73,0.33,0.83}
\definecolor{mediumpurple}{rgb}{0.58,0.44,0.86}
\definecolor{mediumsea}{rgb}{0.24,0.70,0.44}
\definecolor{mediumslate}{rgb}{0.48,0.41,0.93}
\definecolor{mediumspring}{rgb}{0.00,0.98,0.60}
\definecolor{mediumturquoise}{rgb}{0.28,0.82,0.80}
\definecolor{mediumviolet}{rgb}{0.78,0.08,0.52}
\definecolor{midnightblue}{rgb}{0.10,0.10,0.44}
\definecolor{mintcream}{rgb}{0.96,1.00,0.98}
\definecolor{mistyrose}{rgb}{1.00,0.89,0.88}
\definecolor{moccasin}{rgb}{1.00,0.89,0.71}
\definecolor{navajowhite}{rgb}{1.00,0.87,0.68}
\definecolor{navyblue}{rgb}{0.00,0.00,0.50}
\definecolor{navy}{rgb}{0.00,0.00,0.50}
\definecolor{oldlace}{rgb}{0.99,0.96,0.90}
\definecolor{olivedrab}{rgb}{0.42,0.56,0.14}
\definecolor{orange1}{rgb}{1.00,0.65,0.00}
\definecolor{orange2}{rgb}{0.93,0.60,0.00}
\definecolor{orange3}{rgb}{0.80,0.52,0.00}
\definecolor{orange4}{rgb}{0.55,0.35,0.00}
\definecolor{orangered}{rgb}{1.00,0.27,0.00}
\definecolor{orange}{rgb}{1.00,0.65,0.00}
\definecolor{orchid1}{rgb}{1.00,0.51,0.98}
\definecolor{orchid2}{rgb}{0.93,0.48,0.91}
\definecolor{orchid3}{rgb}{0.80,0.41,0.79}
\definecolor{orchid4}{rgb}{0.55,0.28,0.54}
\definecolor{orchid}{rgb}{0.85,0.44,0.84}
\definecolor{palegoldenrod}{rgb}{0.93,0.91,0.67}
\definecolor{palegreen}{rgb}{0.60,0.98,0.60}
\definecolor{paleturquoise}{rgb}{0.69,0.93,0.93}
\definecolor{paleviolet}{rgb}{0.86,0.44,0.58}
\definecolor{papayawhip}{rgb}{1.00,0.94,0.84}
\definecolor{peachpuff}{rgb}{1.00,0.85,0.73}
\definecolor{peru}{rgb}{0.80,0.52,0.25}
\definecolor{pink1}{rgb}{1.00,0.71,0.77}
\definecolor{pink2}{rgb}{0.93,0.66,0.72}
\definecolor{pink3}{rgb}{0.80,0.57,0.62}
\definecolor{pink4}{rgb}{0.55,0.39,0.42}
\definecolor{pink}{rgb}{1.00,0.75,0.80}
\definecolor{plum1}{rgb}{1.00,0.73,1.00}
\definecolor{plum2}{rgb}{0.93,0.68,0.93}
\definecolor{plum3}{rgb}{0.80,0.59,0.80}
\definecolor{plum4}{rgb}{0.55,0.40,0.55}
\definecolor{plum}{rgb}{0.87,0.63,0.87}
\definecolor{powderblue}{rgb}{0.69,0.88,0.90}
\definecolor{purple1}{rgb}{0.61,0.19,1.00}
\definecolor{purple2}{rgb}{0.57,0.17,0.93}
\definecolor{purple3}{rgb}{0.49,0.15,0.80}
\definecolor{purple4}{rgb}{0.33,0.10,0.55}
\definecolor{purple}{rgb}{0.63,0.13,0.94}
\definecolor{red1}{rgb}{1.00,0.00,0.00}
\definecolor{red2}{rgb}{0.93,0.00,0.00}
\definecolor{red3}{rgb}{0.80,0.00,0.00}
\definecolor{red4}{rgb}{0.55,0.00,0.00}
\definecolor{red}{rgb}{1.00,0.00,0.00}
\definecolor{rosybrown}{rgb}{0.74,0.56,0.56}
\definecolor{royalblue}{rgb}{0.25,0.41,0.88}
\definecolor{saddlebrown}{rgb}{0.55,0.27,0.07}
\definecolor{salmon1}{rgb}{1.00,0.55,0.41}
\definecolor{salmon2}{rgb}{0.93,0.51,0.38}
\definecolor{salmon3}{rgb}{0.80,0.44,0.33}
\definecolor{salmon4}{rgb}{0.55,0.30,0.22}
\definecolor{salmon}{rgb}{0.98,0.50,0.45}
\definecolor{sandybrown}{rgb}{0.96,0.64,0.38}
\definecolor{seagreen}{rgb}{0.18,0.55,0.34}
\definecolor{seashell1}{rgb}{1.00,0.96,0.93}
\definecolor{seashell2}{rgb}{0.93,0.90,0.87}
\definecolor{seashell3}{rgb}{0.80,0.77,0.75}
\definecolor{seashell4}{rgb}{0.55,0.53,0.51}
\definecolor{seashell}{rgb}{1.00,0.96,0.93}
\definecolor{sienna1}{rgb}{1.00,0.51,0.28}
\definecolor{sienna2}{rgb}{0.93,0.47,0.26}
\definecolor{sienna3}{rgb}{0.80,0.41,0.22}
\definecolor{sienna4}{rgb}{0.55,0.28,0.15}
\definecolor{sienna}{rgb}{0.63,0.32,0.18}
\definecolor{skyblue}{rgb}{0.53,0.81,0.92}
\definecolor{slateblue}{rgb}{0.42,0.35,0.80}
\definecolor{slategray}{rgb}{0.44,0.50,0.56}
\definecolor{slategrey}{rgb}{0.44,0.50,0.56}
\definecolor{snow1}{rgb}{1.00,0.98,0.98}
\definecolor{snow2}{rgb}{0.93,0.91,0.91}
\definecolor{snow3}{rgb}{0.80,0.79,0.79}
\definecolor{snow4}{rgb}{0.55,0.54,0.54}
\definecolor{snow}{rgb}{1.00,0.98,0.98}
\definecolor{springgreen}{rgb}{0.00,1.00,0.50}
\definecolor{steelblue}{rgb}{0.27,0.51,0.71}
\definecolor{tan1}{rgb}{1.00,0.65,0.31}
\definecolor{tan2}{rgb}{0.93,0.60,0.29}
\definecolor{tan3}{rgb}{0.80,0.52,0.25}
\definecolor{tan4}{rgb}{0.55,0.35,0.17}
\definecolor{tan}{rgb}{0.82,0.71,0.55}
\definecolor{thistle1}{rgb}{1.00,0.88,1.00}
\definecolor{thistle2}{rgb}{0.93,0.82,0.93}
\definecolor{thistle3}{rgb}{0.80,0.71,0.80}
\definecolor{thistle4}{rgb}{0.55,0.48,0.55}
\definecolor{thistle}{rgb}{0.85,0.75,0.85}
\definecolor{tomato1}{rgb}{1.00,0.39,0.28}
\definecolor{tomato2}{rgb}{0.93,0.36,0.26}
\definecolor{tomato3}{rgb}{0.80,0.31,0.22}
\definecolor{tomato4}{rgb}{0.55,0.21,0.15}
\definecolor{tomato}{rgb}{1.00,0.39,0.28}
\definecolor{turquoise1}{rgb}{0.00,0.96,1.00}
\definecolor{turquoise2}{rgb}{0.00,0.90,0.93}
\definecolor{turquoise3}{rgb}{0.00,0.77,0.80}
\definecolor{turquoise4}{rgb}{0.00,0.53,0.55}
\definecolor{turquoise}{rgb}{0.25,0.88,0.82}
\definecolor{violetred}{rgb}{0.82,0.13,0.56}
\definecolor{violet}{rgb}{0.93,0.51,0.93}
\definecolor{wheat1}{rgb}{1.00,0.91,0.73}
\definecolor{wheat2}{rgb}{0.93,0.85,0.68}
\definecolor{wheat3}{rgb}{0.80,0.73,0.59}
\definecolor{wheat4}{rgb}{0.55,0.49,0.40}
\definecolor{wheat}{rgb}{0.96,0.87,0.70}
\definecolor{whitesmoke}{rgb}{0.96,0.96,0.96}
\definecolor{white}{rgb}{1.00,1.00,1.00}
\definecolor{yellow1}{rgb}{1.00,1.00,0.00}
\definecolor{yellow2}{rgb}{0.93,0.93,0.00}
\definecolor{yellow3}{rgb}{0.80,0.80,0.00}
\definecolor{yellow4}{rgb}{0.55,0.55,0.00}
\definecolor{yellowgreen}{rgb}{0.60,0.80,0.20}
\definecolor{yellow}{rgb}{1.00,1.00,0.00}

\title[UV colours from minor mergers at $z=0$]
    {The role of minor mergers in the recent star formation history of early-type galaxies}

\author[Sugata Kaviraj et al.]
{Sugata Kaviraj$^{1}$\thanks{E-mail: skaviraj@astro.ox.ac.uk}, Sebastien Peirani$^{1,2}$, Sadegh Khochfar$^{1,4}$, Joseph Silk$^{1}$ \newauthor and Scott Kay$^{3}$\\
$^{1}$ Department of Physics, University of Oxford, Keble Road,
Oxford, OX1 3RH, UK\\
$^{2}$ Institut d'Astrophysique de Paris, 98bis, bd Arago - 75014 Paris, France\\
$^{3}$ School of Physics and Astronomy, University of Manchester,
Oxford Road, Manchester, M13 9PL, UK\\
$^{4}$ Max Planck Institut f\"ur extraterrestrische Physik, p.o.
box 1312, D-85478 Garching, Germany}

\begin{document}



\pagerange{\pageref{firstpage}--\pageref{lastpage}} \pubyear{2007}

\maketitle

\label{firstpage}


\begin{abstract}
We demonstrate that the large scatter in the ultra-violet (UV)
colours of {\color{black}intermediate-mass} early-type galaxies in
the local Universe and the inferred low-level recent star
formation in these objects can be reproduced by minor mergers in
the standard $\Lambda$CDM cosmology. Numerical simulations of
mergers with mass ratios $\leq$ 1:4, with reasonable assumptions
for the ages, metallicities and dust properties of the merger
progenitors, produce good agreement to the observed UV colours of
the early-type population, if the infalling satellites are assumed
to have (cold) gas fractions $\geq$ 20\%. Early-types that satisfy
$(NUV-r)\lesssim3.8$ are likely to have experienced mergers with
mass ratios between 1:4 and 1:6 within the last $\sim1.5$ Gyrs,
while those that satisfy $3.8<(NUV-r)<5.5$ are consistent with
either recent mergers with mass ratios $\leq$ 1:6 or mergers with
higher mass ratios that occurred more than $\sim 1.5$ Gyrs in the
past. {\color{black}We demonstrate that the early-type
colour-magnitude relations and colour distributions in both the UV
and optical spectral ranges are consistent with the expected
frequency of minor merging activity in the standard $\Lambda$CDM
cosmology at low redshift. We present a strong plausibility
argument for minor mergers to be the principal mechanism behind
the large UV scatter and associated low-level recent star
formation observed in early-type galaxies in the nearby Universe.}
\end{abstract}




\section{Introduction}
The formation histories of early-type galaxies remains a
controversial and often-debated topic in modern astrophysics. The
bulk of the past effort on early-type systems has focussed, almost
exclusively, on their \emph{optical} properties. The optical
colours of the early-type population are predominantly red,
implying that the bulk of the stellar mass in these systems forms
at high redshift \citep[e.g.][]{BLE92}. Furthermore, high
$\alpha$-enhancement ratios in their stellar spectra indicate that
this star formation plausibly takes place over short ($<1$ Gyr)
timescales \citep[e.g.][]{Thomas1999}. However, a drawback of
optical data is its lack of sensitivity to moderate amounts of
\emph{recent star formation} (RSF), within the last Gyr or so. The
optical spectrum remains largely unaffected by the minority of
stellar mass that forms in these systems at low and intermediate
redshift ($z<1$), which makes it difficult to measure early-type
star formation histories (SFHs) over the last 8 billion years
{\color{black} with significant accuracy using optical colours
alone.}

A first step towards probing early-type SFHs over this period is
to quantify their RSF at $z\sim0$. Rest-frame UV photometry
provides an attractive \emph{photometric} indicator of RSF. While
its impact on the optical spectrum is relatively weak (and
virtually undetectable, given typical observational and
theoretical uncertainties), a small mass fraction ($<3$\%) of
young ($<1$ Gyr old) stars strongly affects the rest-frame UV
spectrum shortward of $3000\AA$ \citep{Kaviraj2007d}. Furthermore,
the UV remains largely unaffected by the age-metallicity
degeneracy \citep{Worthey1994} that typically plagues optical
analyses (Kaviraj et al. 2007a), making it an ideal photometric
indicator of RSF.

\begin{figure}
\includegraphics[width=\columnwidth]{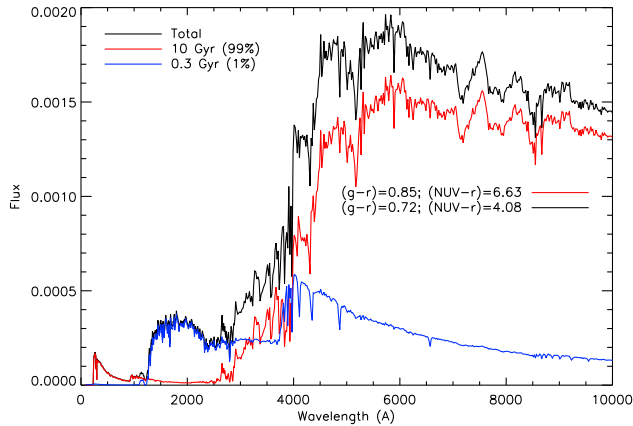}
\caption{The sensitivity of the UV to small mass fractions of
young stellar populations. We construct a model in which an old
(10 Gyr old) population contributes 99\% of the stellar mass
(shown in red), with a 1\% contribution from stars that are $0.3$
Gyrs old (show in blue). The combined spectral energy distribution
(SED) is shown in black. The UV output of the combined SED comes
\emph{purely} from the young population (blue) and that the
\emph{shape} of the optical spectrum - which determines the
optical colours - changes only very slightly. We also indicate the
$(g-r)$ and $(NUV-r)$ colours of the combined population (black)
with those of the purely old population (red). While the ($g-r$)
colour changes by $\sim0.1$ mag from that of a purely old
population, the ($NUV-r$) colour changes by $\sim2.5$ mags!}
\label{fig:rsf_plots}
\end{figure}

{\color{black}In Figure \ref{fig:rsf_plots} we demonstrate the
sensitivity of the UV to small mass fractions of young stellar
populations. We construct a model in which an old (10 Gyr old)
population contributes 99\% of the stellar mass (shown in red),
with a 1\% contribution from stars that are $0.3$ Gyrs old (show
in blue). The combined spectral energy distribution (SED) is shown
in black. The UV output of the combined SED comes \emph{purely}
from the young population (blue) and the \emph{shape} of the
optical spectrum - which determines the optical colours - changes
only very slightly. We also compare the $(g-r)$ and $(NUV-r)$
colours of the combined population (black) and those of the purely
old population (red). While the ($g-r$) colour changes by
$\sim0.1$ mag from that of a purely old population, the ($NUV-r$)
colour changes by $\sim2.5$ mags!}

A new generation of early-type studies that have exploited UV data
from the GALEX space telescope (Martin et al. 2005) have shown
that, in contrast to their optical colour-magnitude relations
(CMRs), nearby ($0<z<0.11$), {\color{black}luminous} ($M(V)<-21$)
early-types show a large spread in their UV colour distribution of
almost 5 mags {\color{black} - a direct consequence of the
sensitivity of the UV to small amounts of recent star formation
that is demonstrated in Figure \ref{fig:rsf_plots} above.}
Following the early work of \citet{Yi2005}, Kaviraj et al. (2007b;
K07 hereafter) have demonstrated that, while a negligible fraction
of the early-type population has photometry consistent with no
star formation within the last 2 Gyrs, \emph{at least} 30\% show
unambiguous signs of RSF, with stellar mass fractions of 1-3\%
forming within the last Gyr, with luminosity-weighted ages of
$\sim300-500$ Myrs.

\begin{figure}
\includegraphics[width=\columnwidth]{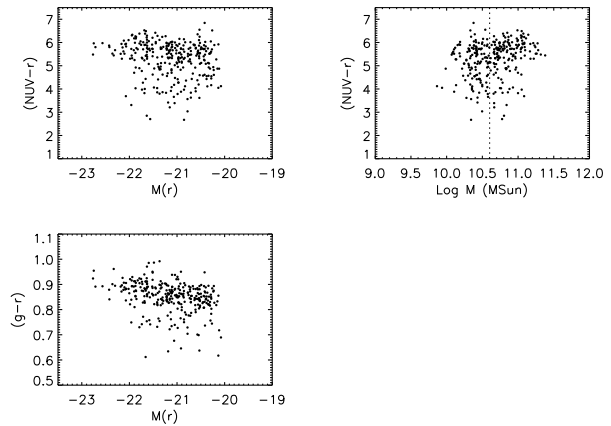}
\caption{TOP LEFT: The ($NUV-r$) colour-magnitude relation of
nearby ($0.05<z<0.06$) early-type galaxies drawn from the SDSS
DR4. Note that the GALEX $NUV$ filter is centred at $\sim2300$
\AA. TOP RIGHT: The ($NUV-r$) colours of SDSS early-types plotted
against their stellar masses, taken from \citet{Gallazzi2005}.
Note that the Gallazzi et al. masses have been released through
the public Garching SDSS catalog which can be found here:
http://www.mpa-garching.mpg.de/SDSS/DR4/. The nominal stellar mass
adopted for the spheroid in our simulations ($4\times 10^{10}$
M$_{\odot}$, see Section 2 below) is indicated using the solid
line. BOTTOM LEFT: The ($g-r$) colour-magnitude relation of nearby
($0.05<z<0.06$) early-type galaxies drawn from the SDSS DR4.}
\label{fig:nuvr_cmr}
\end{figure}

{\color{black} In the {\color{black}top} left-hand panel of Figure
\ref{fig:nuvr_cmr} we show the ($NUV-r$) CMR of nearby
($0.05<z<0.06$) early-type galaxies drawn from the SDSS DR4. Note
that the GALEX $NUV$ filter is centred at $\sim2300$ \AA. The
large spread in the UV colours is in contrast to the small spread
(a few tenths of a mag) in the optical $(g-r)$ colour (bottom
left-hand panel). Following K07, early-type morphology is
established by visually inspecting each individual object using
its SDSS image. The galaxies shown in this figure are selected to
have $r<16.8$, since this is the redshift range within which
visual inspection can be robustly performed using SDSS images (see
Section 2 in K07). Note that the luminosity of a typical
early-type galaxy, in the SDSS $r$-band, lies around $M(r)^* \sim
-21.15$ (see Figure 2 in Bernardi et al. 2003). In the right-hand
panel of this figure we present the ($NUV-r$) colours of the SDSS
early-types plotted against their \emph{stellar masses}, taken
from \citet{Gallazzi2005}. Note that the Gallazzi et al. masses
have been released through the public Garching SDSS catalog which
can be found here: http://www.mpa-garching.mpg.de/SDSS/DR4/. The
nominal stellar mass adopted for the spheroid in our simulations
is $4\times 10^{10}$ M$_{\odot}$ (see also Section 2 below) where
the scatter in the UV CMR becomes broad. This value is indicated
using a {\color{black}dotted} line.}

{\color{black} Note that, while the effects of RSF will be present
in all diagnostics of star formation (including e.g. the optical
$(g-r)$ colour), the sensitivity of a particular indicator depends
on the proportional change due to RSF compared to the typical
uncertainties in that indicator. These uncertainties come both
from measurement errors and from uncertainties in stellar models
that are used to convert spectro-photometric quantities (e.g.
colours) into estimates of physical parameters such as the ages
and mass fractions of young stars. Typical model and observational
errors in optical filters are $\sim0.05$ mags (Sukyoung Yi priv.
comm., see also Yi 2003) and at least 0.02 mags (including
calibration uncertainties) respectively. The total resultant
uncertainty in the optical $(g-r)$ colours is $\sim0.08$ mags
(compared to a spread in this colour of $\sim0.3$ mags from Figure
\ref{fig:nuvr_cmr}). In a similar vein, the typical model and
theoretical errors in the $NUV$ magnitudes are both around 0.2
mags, yielding uncertainties in the $(NUV-r)$ colours of $\sim0.3$
mags (compared to a spread in this colour of almost 5 mags). The
enhanced sensitivity of the $NUV$ to RSF stems from the fact that
the spread in the $NUV$ colours is significantly (possibly an
order of magnitude or more) larger than the uncertainties in the
colours. While RSF does leave an imprint on the optical spectrum,
the overall sensitivity of the $(NUV-r)$ colour is much greater,
making it a much more stringent test of the minor merger
hypothesis than can be performed using optical colours alone.}

While the UV has revealed the unexpected presence of widespread
low-level RSF in the local early-type population, past efforts
have only measured the star formation activity without exploring
the \emph{physical mechanism} for RSF in these galaxies. In this
paper, we explore the potential role of minor mergers in
reproducing the UV colour-magnitude relation of
{\color{black}{\color{black}intermediate-mass}} early-type
galaxies at $z\sim0$. It is worth noting that, although low level
RSF can produce blue UV colours, only stars formed in the last
$\sim$Gyr contribute significantly to the UV flux. Hence, an event
like a minor merger - where one expects an \emph{age profile} in
the recent star formation - may not automatically produce blue UV
colours, even if the supply of cold gas is adequately high to
produce the observed mass fraction of young stars. In this study,
we combine the results of numerical simulations of minor mergers
with standard stellar models to probe the photometric properties
of such events and test consistency with the observations, given
reasonable assumptions for the properties of the merger
progenitors and the predicted frequency of merging activity in the
$\Lambda$CDM paradigm.


\section{Simulations}
{\color{black}In this section we describe the numerical
methodology used to study minor mergers between a typical
elliptical galaxy and a satellite, where the mass ratio of the
system is between 1:4 and 1:10}. A more complete description will
be provided in Peirani et al. (in prep). {\color{black}Note that
the nominal stellar mass of the spheroid used to describe the
photometric properties of the simulations in Section 3 matches the
region of the $(NUV-r)$ vs. mass plot (see Figure
\ref{fig:nuvr_cmr} above) where the scatter is broadest.}


\subsection{Initial conditions}
The elliptical is modelled using spherical dark matter (DM) halo
and stellar components, with a total mass of $10^{12} M_{\odot}$.
Following \citet{Jiang2007}, stars contribute 4\% of this value.
The DM halo follows a Hernquist profile \citep{Hernquist1990},
with parameters chosen so that the inner part coincides with an
NFW profile \citep{NFW1996} with a virial radius $r_{200}=206$ kpc
and a concentration parameter $C=10$, as indicated by previous
cosmological N-body simulations \citep{Dolag2004}. A Hernquist
profile reproduces the `de Vaucouleurs' $R^{1/4}$ surface
brightness profiles of typical elliptical galaxies. The effective
radius of the projected brightness is $r_e=4.3$ kpc.

The satellite is constructed using a spherical DM halo (with a
Hernquist profile) which contains a disk containing stars and gas
but no bulge. The mass of the disk represents 5\% of the total
mass, with a gas fraction of either 20\% or 40\%. The gas
fractions are consistent with the observed values from the SDSS
\citep{Kannappan2004}. Satellites are created following
\citet{Springel2005a} and their rotation curves satisfy the
baryonic Tully-Fisher relation. In all simulations satellites are
put on prograde parabolic orbits \citep{Khochfar2006b}, with a
pericentric distance $R_{p}=8$ kpc, and initial separations of
$100$ kpc.

\subsection{Numerical method}
The simulations are performed using the public GADGET2 code
\citep{Springel2005b} with added prescriptions for cooling, star
formation and feedback from Type Ia and II supernovae (SN).
Approximately 380,000 particles are used for each experiment and
the particle masses and gravitational softening lengths
($\epsilon$) involved are summarized in the table below.

\small
\begin{center}
\begin{tabular}{c|c|c|c|c}
\hline
 & DM & gas & star (disk) & star (E)\\
\hline
Mass ($10^{5} M_\odot$) & 30.3 & 4.5 & 4.5 & 13.5\\
\hline
$\epsilon$ (kpc) & 0.4 & 0.5 & 0.5 & 0.4\\
\hline
\end{tabular}
\end{center}
\normalsize

The cooling and star formation (SF) recipes follow the
prescriptions of \citet{Thomas1992} and \citet{Katz1996}
respectively. Gas particles with $T>10^4$K cool at constant
density (with the assumption of solar metallicity) for the
duration of each timestep. Gas particles with $T< 2\times 10^4 K$,
number density $n > 0.1\, cm^{-3}$, overdensity $\delta
\rho_{gas}> 100$ and ${\bf \nabla . \upsilon} <0$ form stars
according to the standard star formation prescription: $d\rho_*/dt
= c_* \rho_{gas}/t_{dyn}$, where $\rho_*$ refers to the stellar
density, $t_{dyn}$ is the dynamical timescale of the gas and $c_*$
is the SF efficiency. Instead of creating new (lighter) star
particles, we implement the SF prescription in a probabilistic
fashion. Assuming a constant dynamical time across the timestep,
the fractional change in stellar density, $\Delta \rho_*/\rho_* =
1-\exp(-c_* \Delta t/t_{\rm dyn})$. For each gas particle, we draw
a random number ($r$) between 0 and 1 and convert it to a star if
$r<\Delta \rho_*/\rho_*$.

The energy injection into the inter-stellar medium (ISM) from SN,
which regulates the star formation rate (SFR), is modelled
following the approach of Durier \& de Freitas Pacheco (2007, in
prep.). Instead of assuming `instantaneous' energy injection, we
include the effective lifetime of SN progenitors using the rate of
energy injection $H_{SN}$. For this, we consider stellar lifetimes
in the mass ranges $0.8\,M_\odot<m<8.0\,M_\odot$ and
$8.0\,M_\odot<m<80.0\,M_\odot$ for Type Ia and Type II progenitors
respectively. Using a Salpeter initial mass function for Type II
SN gives:

\begin{equation}
H_{SN_{II}}=2.5\times10^{-18}\Big(\frac{m_*}{M_\odot}\Big)E_{SN}\Big(\frac{1300}{\tau(\textnormal{Myr})-3}\Big)^{0.24}
\textnormal{erg.s$^{-1}$}
\end{equation}

\noindent where $E_{SN}=10^{51}$ erg, $m_*$ is the mass of the
star particle and $3.53 <\tau <29$ Myr. For Type Ia SN, the
heating is delayed, since they appear $t_0=0.8-1.0$ Gyr after the
onset of star formation. Following \citet{deFreitasPacheco1998},
the probability of one event in a timescale $\tau$ after the onset
of star formation is given by:

\begin{equation}
H_{SN_{I_a}}=4.8\times10^{-20}\Big(\frac{m_*}{M_\odot}\Big)E_{SN}\Big(\frac{t_0}{\tau}\Big)^{3/2}
\textnormal{erg.s$^{-1}$}
\end{equation}

Eqns (1) and (2) are used to compute the energy released ($E_i$)
by SN derived from a star particle $i$, and a fraction $\gamma$ of
this energy is deposited in the j$^{th}$ neighbour gas particle by
applying a radial kick to its velocity with a magnitude $\Delta
v_j = \sqrt{(2w_j\gamma E_i/m_j)}$, where $w_j$ is the weighting
based on the smoothing kernel and $m_j$ is the mass of gas
particle j. We note that all gas neighbours are located in a
sphere of radius $R_{SN}$, centered on the SN progenitor, to avoid
spurious injection of energy outside the SN's region of influence.

In the simulations presented in this study, we use $\gamma=0.1$,
$R_{SN}=0.4$ kpc and vary $c_*$ in the range $0.01<c_*<0.1$. When
satellites are isolated, these parameters lead to a quasi-constant
SFR between 0.2 and $3.0\,M_\odot$yr$^{-1}$, depending on the mass
and initial gas fraction of the satellite. These SFRs are in good
agreement with those for low-mass objects found in previous
simulations of isolated galaxies \citep[e.g.][]{Stinson2006}.

\begin{figure}
\rotatebox{0}{\includegraphics[width=\columnwidth]{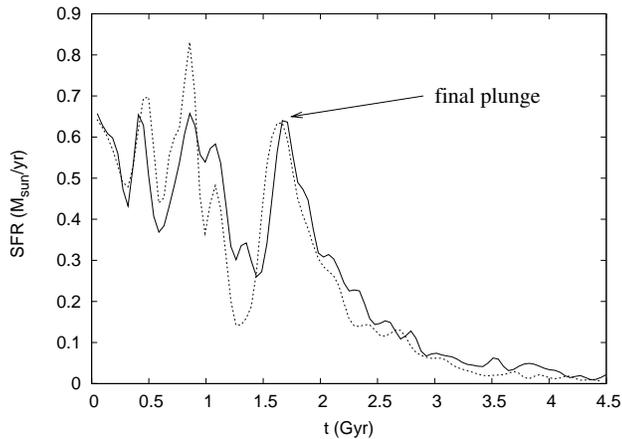}}
\caption{The star formation rate of a minor major with mass ratio
1:6, where the gas fraction of the satellite is 20\% and
$c_*=0.05$. The dashed line represents the same simulation with 10
times more particles. At the `final plunge' the satellite
disappears into the parent elliptical, so that images (e.g.
through an SDSS $r$-band filter) would indicate a single
spheroidal object.} \label{fig:sfr}
\end{figure}

Finally, we have checked that increasing the resolution of the
simulations (by a factor of 10) does not affect the star formation
history or the derived colours so that the conclusions presented
in this work are robust. A typical example of the SFH is shown in
Figure \ref{fig:sfr}.

\section{The case for minor mergers}
We begin by exploring the predicted photometry from an ensemble of
minor merger simulations (described in Section 2) with the
assumptions that (a) both merger progenitors have solar
metallicity, (b) the elliptical and satellite have
luminosity-weighted ages of 9 and 5 Gyrs respectively and (c) the
dust extinction due to the ISM in the remnant is given by
$E_{B-V}^{ISM}\sim0.05$.

{\color{black}We briefly describe the motivation for modelling the
underlying population of the spheroid using a 9 Gyr old simple
stellar population (SSP). It is well established that the bulk of
the stellar populations in elliptical galaxies are uniformly old.
This is demonstrated by the fact that (a) they show red optical
colours with small scatter \citep[e.g.][]{BLE92,Stanford98} and
(b) they exhibit high alpha-enhancement (e.g. [Mg/Fe]) ratios,
which implies that the star formation took place on timescales
shorter than the typical onset timescales of Type Ia supernovae
i.e. $<1$ Gyrs \citep[see e.g.][]{Thomas1999}. Hence the
underlying population can be approximated by an old SSP. The
particular choice of 9 Gyrs is motivated by the fact that Bernardi
et al. (2003), who recently studied the SDSS elliptical galaxy
population in the nearby Universe, were able to fit their optical
colour-magnitude relations with an SSP with an age of 9 Gyrs. It
is worth noting that \emph{optical} colour evolution virtually
stops after $\sim6$ Gyrs \citep{Yi2003}. This means that a 6 Gyr
old SSP looks very similar to an older stellar population e.g. 9
Gyrs. In other words we could replace the 9 Gyr SSP with a 6 or 10
Gyr SSP and our results will not change. Note that similar
techniques (i.e. using an old SSP to represent the underlying
stellar population of ellipticals) have been frequently used by
previous studies. e.g. \citet{Trager2000b} and
\citet{Ferreras2000}. Finally, stellar populations that are older
than $\sim2$ Gyrs do not contribute to the UV (see the bottom
panel of Figure 7 in K07). As a result, the \emph{underlying}
population of an elliptical galaxy will not contribute to the UV
at all.}

The estimate for the ISM extinction is an average value for local
early-types, derived by K07 from parameter estimation using GALEX
(UV) and SDSS (optical) photometry. Since the youngest stars are
expected to reside in molecular clouds (MCs) which have short
lifetimes of a few tens of Myrs
\citep[e.g][]{Blitz1980,Hartmann2001} and dust extinction several
times larger than that due to the ISM alone
\citep[e.g.][]{Charlot2000}, we explore MC lifetimes in the range
0-50 Myrs and MC extinctions in the range $0.05<E_{B-V}^{MC}<0.5$.
Thus, when the synthetic photometry from the simulations is
constructed, stars with ages less than the MC lifetime in question
are subject to the prescribed MC extinction. The stellar models
used in this study are described in \citet{Yi2003}.

\begin{figure}
\begin{center}
\includegraphics[width=3.5in]{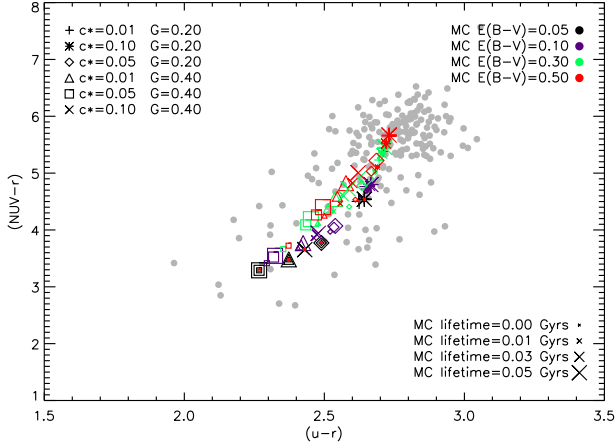}
\caption{Predicted photometry from an ensemble of minor merger
simulations with mass ratios of 1:10. Both progenitors are assumed
to have solar metallicity with the elliptical and the satellite
having ages of 9 and 5 Gyrs respectively. The ISM dust extinction
in the remnant is given by $E_{B-V}^{ISM}\sim0.05$. Symbol types
represent merger configurations - the star formation efficiency
($c*$) and gas fraction ($G$) of the accreted satellites are shown
in the top left legend. MC extinctions are shown using colours,
while symbol sizes represent MC lifetimes. Grey dots represent the
observed colours of luminous ($L^*$ or above) early-type galaxies
in the redshift range $0.05<z<0.06$. Note that typical
uncertainties in the observed colours are 0.2 mag and that the
synthetic photometry has been redshifted to $z=0.065$ for a direct
comparison.} \label{fig:bcprops1_10}
\end{center}
\end{figure}

In Figure \ref{fig:bcprops1_10} we present the synthetic
photometry from various configurations for mergers with mass
ratios of 1:10. The remnant is `observed' at the point where the
satellite finally disappears into the parent elliptical, so that
images (e.g. through an SDSS $r$-band filter) would indicate a
single spheroidal object. The colours shown are therefore the
\emph{bluest} possible for each scenario where the system appears
to be one object. Star formation declines after this `final
plunge' (see Figure \ref{fig:sfr}) and the remnant reddens, in the
$(NUV-r)$ colour, by $\sim0.8$ mag/Gyr for mass ratios of 1:4 and
1:6 and by $\sim0.5$ mag/Gyr for a mass ratio of 1:10.

We find that, with reasonable MC properties - e.g. lifetimes
$\geq30$ Myrs (red and green colours) and high dust extinctions
(large symbols) - 1:10 mergers can reproduce most but not all of
the scatter in the observed UV colours. In particular, the
\emph{bluest} UV colours ($NUV-r\lesssim4$) cannot be accounted
for by 1:10 mergers because the RSF induced is inadequate, even
when the satellite has a high gas fraction ($\sim40$\%). The
observed early-type UV colour-magnitude relation (CMR; shown using
filled grey circles in Figure \ref{fig:bcprops1_10}) is restricted
to the redshift range $0.05<z<0.06$ (to ensure robust
morphological classification from SDSS images) and $r<16.2$ (which
corresponds to galaxies more luminous than $L^*$ at this
redshift).

\begin{figure}
\begin{center}
$\begin{array}{c}
\includegraphics[width=0.5\textwidth]{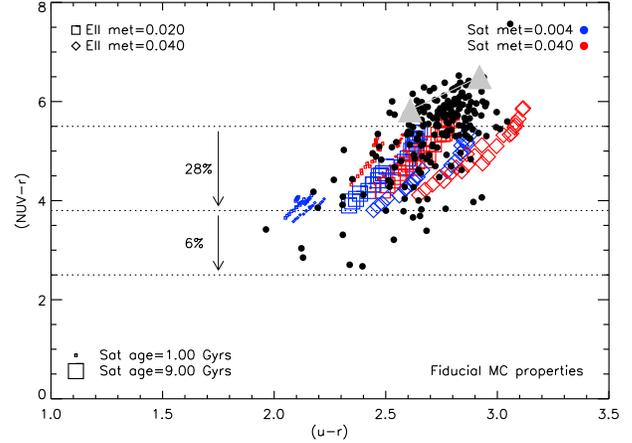}
\end{array}$
\caption{TOP: Predicted photometry from an ensemble of minor
mergers with mass ratios of 1:10. We assume a spread in the age
(1-9 Gyrs; shown using symbol sizes) and metallicity
($0.2Z_{\odot}-2Z_{\odot}$; shown using colours) of the satellite.
The observed colours of the local,
{\color{black}{\color{black}intermediate-mass}} ($L^*$ or above)
early-type population is shown by the black dots. Large grey
triangles indicate the positions of simple stellar populations
with half-solar and solar metallicities that form at $z=3$.}
\label{fig:age_met_spread}
\end{center}
\end{figure}

Recalling that Figure \ref{fig:bcprops1_10} assumes solar
metallicity and a single age for the satellite (5 Gyrs), we now
explore a wider parameter space where we vary the metallicity of
the satellite in the range $0.2Z_{\odot}$ to $2Z_{\odot}$ and its
luminosity-weighted age in the range 1-9 Gyrs. The analysis is
restricted to realistic MC properties - lifetimes $\gtrsim$ 30
Myrs and extinction $\gtrsim$ 0.3. We present these results in
Figure \ref{fig:age_met_spread}. While the predicted photometry
falls in the centre of the locus of observed galaxies in Figure
\ref{fig:bcprops1_10}, we find that a reasonable spread in the age
and metallicity of the satellite can reproduce the `horizontal'
spread in the $(u-r)$ colours. However, such a spread in age and
metallicity cannot mimic low MC lifetimes/extinctions i.e. if we
restrict our analysis to realistic MC properties only, low
satellite ages and metallicities remain unable to reproduce the
bluest UV colours. We also indicate, in Figure
\ref{fig:age_met_spread}, the early-type galaxy fractions bluer
than $(NUV-r)<3.8$ (the colour limit of the 1:10 mergers with
realistic MC properties) and in the colour range
$3.8<NUV-r\leq5.5$. K07 estimated that, within theoretical and
observational uncertainties, galaxies bluer than $(NUV-r)\sim5.5$
are {\color{black}very} likely to have had some RSF (within the
last $\sim$Gyr), while galaxies with $(NUV-r)\geq5.5$ can be
consistent with both RSF and purely old ($>2$ Gyrs old) stellar
populations.

\begin{figure}
\begin{center}
$\begin{array}{c}
\includegraphics[width=0.5\textwidth]{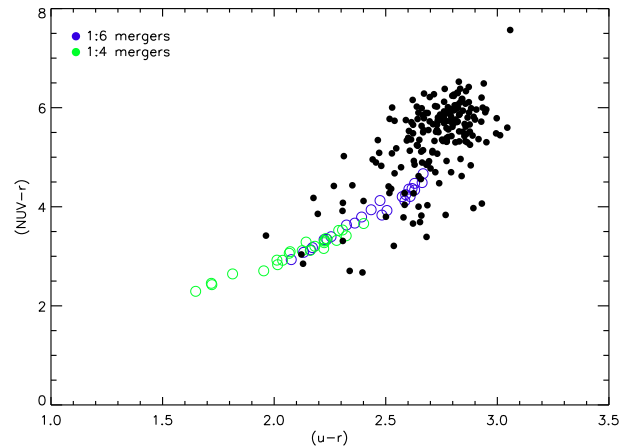}\\
\end{array}$
\caption{Predicted photometry from the same ensemble of minor
merger simulations as in Figure \ref{fig:bcprops1_10} but with
mass ratios of 1:6 (blue) and 1:4 (red). Both progenitors are
assumed to have solar metallicity with the elliptical and the
satellite having ages of 9 and 5 Gyrs respectively. The ISM dust
extinction in the remnant is given by $E_{B-V}^{ISM}\sim0.05$. We
only show configurations with realistic MC properties - lifetimes
$\gtrsim$ 30 Myrs and extinction $\gtrsim$ 0.3}
\label{fig:bcprops1_6and1_4}
\end{center}
\end{figure}

We now repeat the analysis with mergers that have mass ratios of
1:6 and 1:4. Figure \ref{fig:bcprops1_6and1_4} indicates that the
blue end of the early-type UV colours, which is inconsistent with
1:10 mergers, can indeed be reproduced by 1:6 and 1:4 mergers with
realistic assumptions for MC properties.

\section{Merger statistics in $\Lambda$CDM and reproduction of the observed early-type colours}
\begin{figure}
\begin{center}
\includegraphics[width=3.5in]{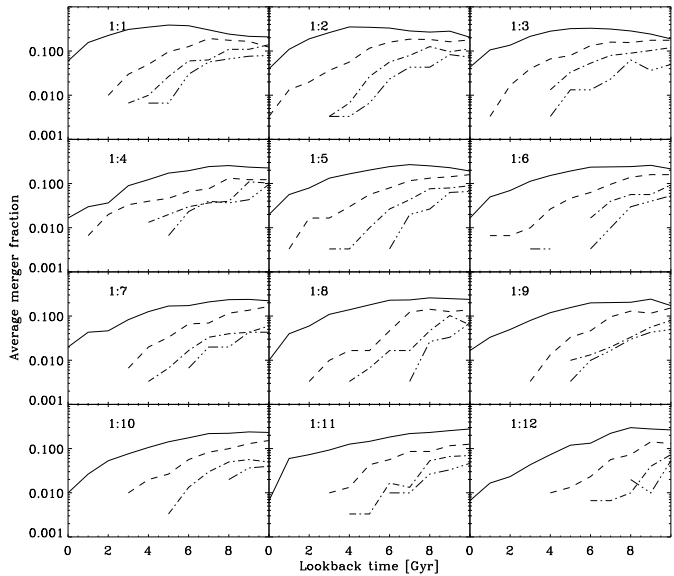}
\caption{{\color{black}The average fraction of spheroidal galaxies
(in dark matter halos of mass $10^{12}$ M$_{\odot}$ or above) that
are predicted to have had one (solid), two (dashed), three
(dot-dashed) and four (triple dotted) 1:X mergers in a given
look-back time in $\Lambda$CDM. The value of `X' is indicated in
each panel and the dispersions in the fractions are $\sim20$\%.
Fractions are calculated by dividing the number of mergers that
occurred in look-back time ($t$) bins in the histories of
spheroids by the number of such spheroids at $z=0$. For example,
the point at $t=0$ represents the merger fraction within a
look-back time of 0 and 1 Gyr, while the point at $t=1$ represents
the merger fraction within a look back time of 1 and 2 Gyrs and so
on. The merger fraction increases with look-back time because the
merger rate in the Universe increases at higher redshift
\citep[see also e.g.][]{Gottlober2001,LeFevre2000}. Note that the
merger fraction increases monotonically until the merger rate in
the Universe peaks and then drops off.}}
\label{fig:merger_statistics}
\end{center}
\end{figure}

While the UV colours of
{\color{black}{\color{black}intermediate-mass}} early-type
galaxies appear consistent with minor mergers, the reproducibility
of the entire UV CMR depends on the frequency of merging activity
at low redshift. In Figure \ref{fig:merger_statistics} we present
the average fraction of early-type galaxies {\color{black}(in dark
matter halos of mass $10^{12}$ M$_{\odot}$ or above)} that are
predicted to have had one (solid), two (dashed), three
(dot-dashed) and four (triple dotted) 1:X mergers in a given
look-back time in $\Lambda$CDM. The value of `X' is indicated in
each panel. Merger trees are generated using the semi-analytical
model of Khochfar and Burkert (2003, 2005) and
\citet{Khochfar2006} and morphology is traced using stellar
bulge:total ($B/T$) ratios - early-type galaxies are assumed to
have $B/T>0.7$.

{\color{black}The merger fractions are calculated by dividing the
number of mergers that occurred in look-back time ($t$) bins in
the histories of spheroids by the number of such spheroids at
$z=0$. For example, the point at $t=0$ represents the merger
fraction within look-back times of 0 and 1 Gyr, while the point at
$t=1$ represents the merger fraction within look back times of 1
and 2 Gyrs and so on. The merger fraction increases with look-back
time because the merger rate in the Universe increases at higher
redshift \citep[see also e.g.][]{Gottlober2001,LeFevre2000}. Note
that the merger fraction increases monotonically until the merger
rate in the Universe peaks and then drops off.}

{\color{black}Before comparing the photometric predictions from
our model machinery to the observations, we briefly note that,
while major mergers (which have mass ratios between 1:1 and 1:3)
might also be expected to contribute to the scatter in the UV CMR,
a study of the GALEX $NUV$ photometry of a sample of
\emph{ongoing} major mergers, identified by \citet{McIntosh2007}
from the SDSS, indicates that major merger progenitors lie on the
broad UV red sequence ($NUV-r>4.7$). While a caveat here is that
the McIntosh et al. sample is small (compared to the SDSS galaxy
population and especially after cross-matching with GALEX), it is
reasonable to conclude that major mergers will contribute only to
the broadness of UV `red sequence' ($NUV-r>4.7$), not to the full
extent of the UV scatter and certainly not to the blue end of the
UV CMR.}

\begin{figure}
\begin{center}
$\begin{array}{c}
\includegraphics[width=0.5\textwidth]{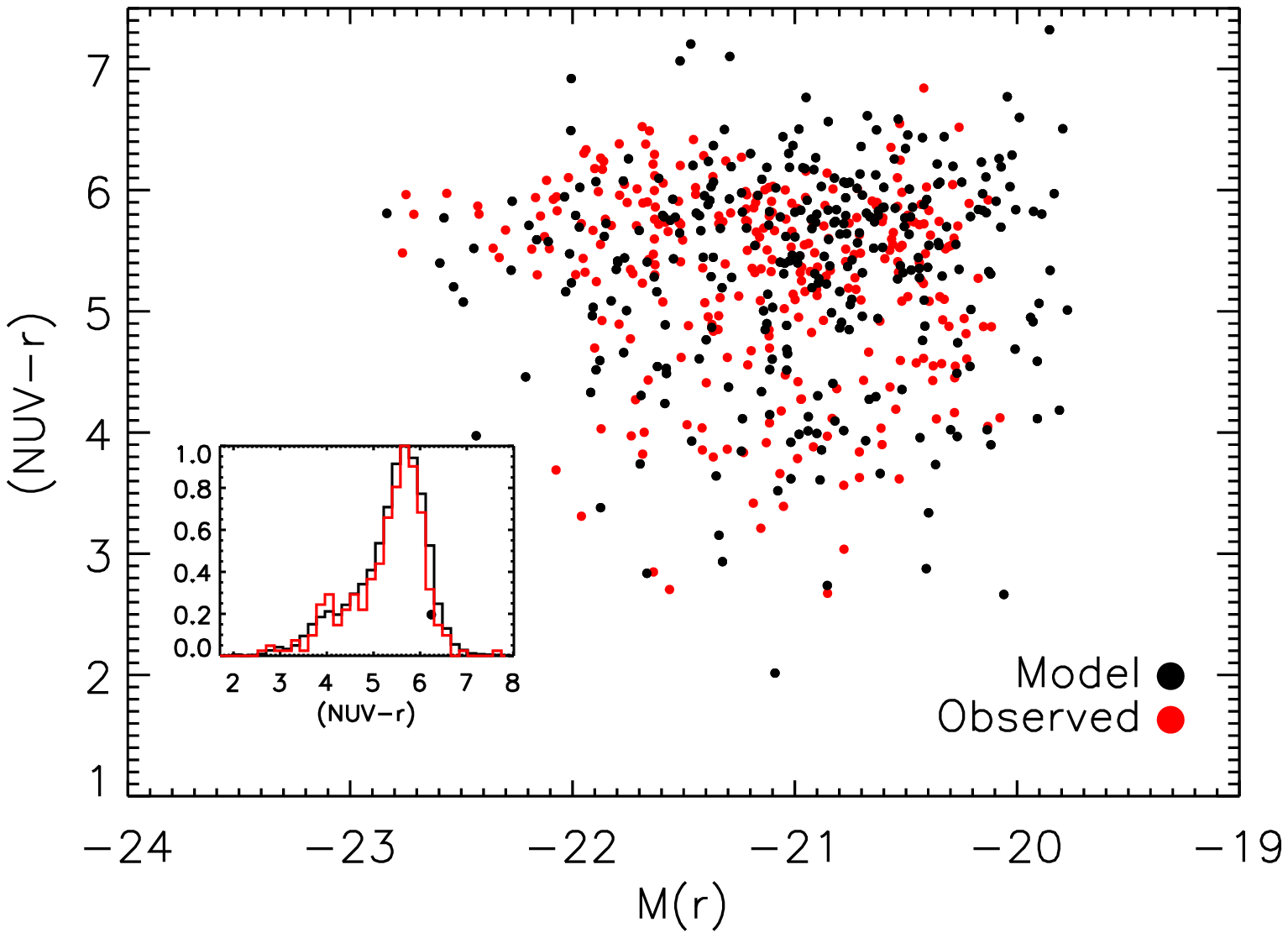}\\
\includegraphics[width=0.5\textwidth]{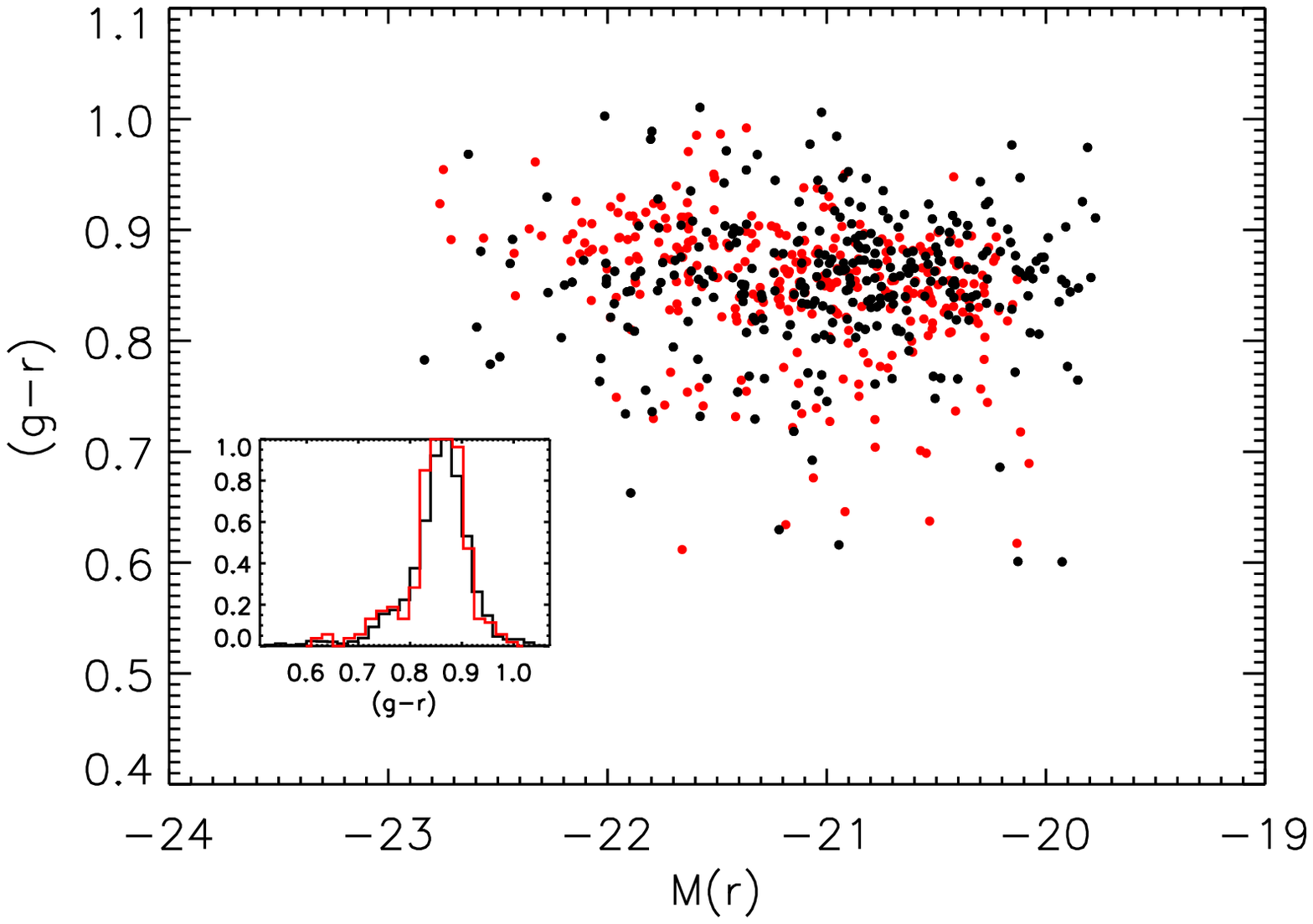}
\end{array}$
\caption{TOP: Comparison of the simulated $(NUV-r)$ colours,
generated by convolving the photometric predictions from minor
merger simulations with the predicted minor merger activity in
LCDM (black), with the observed $(NUV-r)$ colours of the
low-redshift early-type population. Colour histograms (normalised
to 1) are shown in the inset. BOTTOM: The corresponding plot for
the optical $(g-r)$ colour.} \label{fig:mcsims}
\end{center}
\end{figure}

{\color{black}We now check if the predicted LCDM minor merger
activity (i.e. events with mass ratios between 1:4 and 1:10),
convolved with the predictions from the basic set of numerical
simulations described above, is able to simultaneously reproduce
the UV and optical CMRs of the low-redshift early-type population.
Using the SDSS luminosity function of low-redshift early-type
galaxies extracted in K07, we construct a large Monte-Carlo (MC)
ensemble of 50,000 simulated objects and compare the synthetic
CMRs from this ensemble with the observed early-type CMRs in both
the $(NUV-r)$ and $(g-r)$ colours. We assume (a) realistic MC
properties i.e. random MC lifetimes between 30 and 50 Myrs and
$0.3<E_{B-V}^{MC}<0.5$ (b) a gaussian distribution in the
SSP-weighted satellite ages, which peaks at 5 Gyrs, with a width
of 2 Gyrs and (c) a uniform distribution in the satellite
metallicities in the range $0.2Z_{\odot}-2Z_{\odot}$. We only
consider merger events within the last 2.5 Gyrs and note that
including larger time windows does not alter our results because
the UV flux decays rapidly after 2 Gyrs anyway.

In Figure \ref{fig:mcsims} we draw a random subset, equal to the
number of observed galaxies in our comparison redshift range
($0.05<z<0.06$), from the parent MC ensemble of 50,000 objects and
compare their $(NUV-r)$ and $(g-r)$ colours to that of the
observed early-type population. A comparison between the colour
histograms of the parent MC ensemble and the observed early-type
population is also shown in the inset to each figure. Note that
both histograms are normalised to 1. We find that, given the
assumptions listed above, there is good quantitative agreement
between the synthetic and observed CMRs and colour distributions
in both UV and optical colours. In other words, the predicted
minor merger activity in the standard model is able to reproduce
the UV/optical properties of the early-type population (in
particular the distribution scatter to blue UV colours) with a
satisfactory level of accuracy. This, in turn, provides a very
strong plausibility argument for minor merging driving the recent
star formation observed in early-type galaxies at low redshift.}

{\color{black}\subsection{A note about observational signatures of
merging} It is worth noting here that the arguments presented
above imply that an appreciable number of early-type systems in
the nearby Universe should either be in a `closepair' or
`pre-merger' system with a satellite or exhibit morphologies that
are consistent with recent merging events.

We should mention first that several observational studies have
looked at closepairs from a variety of surveys across a range of
redshifts \citep[e.g.][]{Patton2000,LeFevre2000}. While such
studies are very useful, a potential complication is that the
detection of closepairs relies on a range of criteria.
\citet{Patton2000}, for example, tag pairs of galaxies as
`pre-mergers' where the tranverse separation on the sky is less
than 20$h^{-1}$ kpc and the relative velocities are less than 500
kms$^{-1}$. The number of closepairs naturally depends on the
criteria being used. For example, making the criteria stricter
increases the likelihood that each candidate system is truly a
`pre-merger'. However it also results in fewer pre-mergers being
found in total. An added problem is that computing relative
velocities requires spectroscopic redshifts. As a result smaller,
fainter galaxies at the magnitude limit of typical spectroscopic
surveys will not have redshifts and therefore minor mergers cannot
be efficiently identified (since only the larger progenitor has a
redshift). Hence the lack of large numbers of minor merger
closepairs in typical observational studies is not a good
indicator of the true frequency of such events in the nearby
Universe.

A better alternative is to look for \emph{post-mergers} i.e.
objects that show tell-tale signatures of recent merger activity.
An added advantage of studying post-merger systems is that one is
sure that the merger has actually taken place, whereas with
pre-merger systems it is just a guess (albeit a well-educated
one!). While a large observational survey like the SDSS could be
the perfect test-bed for performing such a study, the typical 50
second exposure SDSS images are not deep enough to detect faint
morphological disturbances from minor mergers. However, van Dokkum
(2005) have recently used very deep optical photometry to show
that over 70\% of early-types on the optical \emph{red} sequence
show morphological signatures of merging. The morphological
features, e.g. fans, tails, shells, are faint and red to surface
brightness limits of $\mu\sim28$ mag arcsec$^{-2}$) and consistent
with minor mergers (where the induced star formation is at a very
low-level). These features do not appear in the corresponding SDSS
images and an analysis of the UV magnitudes of these red mergers
(using GALEX) indicates that their UV CMR is similarly broad to
what has been found for the general SDSS early-type population in
K07 (Kaviraj and van Dokkum, in prep).}


\section{Discussion and summary}
We have compared the UV colours of nearby ($0.05<z<0.06$)
early-type galaxies with synthetic photometry derived from
numerical simulations of minor mergers, with reasonable
assumptions for the ages, metallicities and dust properties of the
merger progenitors. {\color{black}\emph{Observational} estimates
for satellite gas fractions have been taken from
\citet{Kannappan2004} and minor merger simulations have been
performed using these gas fractions. We have then appealed to the
merger statistics in the standard $\Lambda$CDM paradigm to check
whether the minor merger activity could plausibly drive the
scatter in the UV CMR at low redshift.

We have found that the bluest end of the early-type UV CMR
($NUV-r<3.8$) is consistent with mergers that have mass ratios
between 1:4 and 1:6 (and cannot be reproduced by events with mass
ratios less than or equal to 1:10), assuming that the infalling
satellites have gas fractions around $\sim20$\% or higher,
{\color{black}which are consistent with the observationally
constrained gas fractions from \citet{Kannappan2004}}. Early-types
with intermediate UV colours ($3.8<NUV-r<5.5$) are consistent with
either recent minor mergers with mass ratios less than 1:6 or
mergers with higher mass ratios more than $\sim1$ Gyr in the past.
{\color{black}Major mergers are likely only to contribute to the
broadness of the UV red sequence and not to the blue scatter in
the UV CMR.} {\color{black}Furthermore, we have demonstrated that
the predicted minor merger activity in the standard model,
convolved with photometric predictions from our fiducial set of
numerical simulations, is able to simultaneously reproduce the UV
and optical CMRs and colour distributions of the low-redshift
early-type population, in particular the large scatter to blue
colours. This, in turn, provides a strong plausibility argument
for minor mergers being responsible for the RSF in early-type
galaxies in the nearby Universe.}

We note here that our study does not utilise a full-blown
semi-analytical model for two important reasons. Firstly, the
amount of gas available at late times in such a model depends on
the baryonic recipes implemented within it. Different models can
predict different gas fractions in satellites at late times.
Secondly, in current semi-analytical models the merger event is
modelled `instantaneously' i.e. the full age profile of stars
formed in the merger is not taken into account. The correct way to
implement minor merger events in a semi-analytical model is to
apply star formation histories from simulations such as those
presented in this paper to merger events in the model. A full
analysis with such an implementation will be provided in a
forthcoming paper (Khochfar et al., in prep). In this paper we
have appealed \emph{only} to the merger fractions predicted by
LCDM, which are robust. Finally, the primary reason for using the
Khochfar and Burkert model is that it uses the Extended Press
Schechter (EPS) formalism to generate merger trees, resulting in
infinite mass resolution which is important because we are
specifically looking at minor mergers involving small (satellite)
objects.

{\color{black}Finally, we note some possible caveats to the
analysis presented here. Firstly, although it is reasonable to
infer relatively high gas fractions ($>20$\%) for infalling
satellites based \citet{Kannappan2004}, it is worth noting that
the gas fractions are derived from calibrations calculated by
combining SDSS photometry and HI measurements. Confirmation of
these gas fractions (especially in satellite galaxies) requires
larger and deeper surveys that yield gas mass measurements, which
are unavailable right now but may be possible using future
instruments such as Herschel. Secondly, we note that although
cooling from hot gas halos could potentially contribute to the
supply of cold gas that then drives star formation, this channel
is unlikely in galaxies in the mass range studied here.
\citet{Dekel2006} show that, at late epochs ($z<2$), the gas in
halos above a critical shock-heating mass ($10^{12}$M$_{\odot}$;
consistent with the galaxies considered in this study) is heated
by a virial shock, leading to long cooling timescales that
effectively shut down the gas supply and subsequent star
formation. Furthermore, unchecked accretion from hot gas halos
would result in early-types becoming too massive and too blue
\citep{Benson2003}. Since there is no reason to believe that that
$z\sim0$ is a preferential epoch for gas cooling, evidence from
the luminosity function of massive galaxies renders star formation
from cooling flows very unlikely. This, combined with the
abundance of minor merger features in the \citet{VD2005} study,
strongly indicates that the recent star formation is driven not by
gas cooling but by merging activity.}

{\color{black}While this study provides a strong plausibility
argument for minor mergers being the principal mechanism behind
the large UV scatter and associated low-level recent star
formation in early-type galaxies, similar studies are required at
intermediate redshifts ($0.5<z<1$) to check whether the evolution
of this scatter is consistent with the $\Lambda$CDM merger
statistics. In addition, deep/high-resolution imaging i.e. with
the Hubble Space Telescope are required to confirm the
\emph{coincidence} of morphological signatures produced by merging
with UV excess in these galaxies. Current galaxy surveys such as
COSMOS (Scoville et al. 2007) make such analyses possible and
results from such studies will be presented in a forthcoming
paper.}


\section*{Acknowledgements}
We warmly thank the anonymous referee for an insightful review
that considerably improved the quality of the original manuscript.
Sugata Kaviraj acknowledges a Leverhulme Early-Career Fellowship
(till Oct 2008), a Research Fellowship from the Royal Commission
for the Exhibition of 1851 (from Oct 2008), a Beecroft Fellowship
from the BIPAC Institute and a Senior Research Fellowship from
Worcester College, Oxford. S. Peirani acknowledges support from
ANR. Finally, we thank J. A. de Freitas Pacheco, F. Durier, I
Ferreras, S. K. Yi and A. Pipino for stimulating conversations.


\nocite{Kaviraj2006} \nocite{Martin2005} \nocite{Khochfar2003}
\nocite{Khochfar2005} \nocite{Kaviraj2007a} \nocite{Kaviraj2007b}
\nocite{Bernardi2003c} \nocite{Scoville2007}


\bibliographystyle{mn2e}
\bibliography{references}


\end{document}